\documentclass[12pt]{article}
\usepackage[round]{natbib}   
\usepackage[hidelinks]{hyperref}
\usepackage{amsfonts}
\usepackage{dsfont}

\usepackage{bm}
\usepackage{color}
\usepackage{nccmath}
\usepackage{algorithm}
\usepackage{algpseudocode}
\usepackage{comment}
\usepackage{caption}
\newtheorem{example}{Example}

\usepackage{amsmath}  
\usepackage{graphicx} 
\usepackage{natbib}   
\usepackage{setspace}
\usepackage{geometry}
\title{Active Learning of Computer Experiment with both Quantitative and Qualitative Inputs}
\author{
    Anita Shahrokhian\thanks{Department of Mathematics and Statistics, Queen's University, Kingston, ON.}, 
        Xinwei Deng\thanks{Department of Statistics, Virginia Tech, Blacksburg, VA.}, 
    and C. Devon Lin\footnotemark[1]
}
\date{}

\begin{document}
\setstretch{1.5}

\maketitle

\begin{quote}{\bf Abstract:}
Computer experiments refer to the study of real systems using complex simulation models. They have
been widely used as alternatives to physical experiments. Design and analysis of computer experiments
have attracted great attention in past three decades. The bulk of the work, however, often focus on
experiments with only quantitative inputs. In recent years, research on design and analysis for computer
experiments have gain momentum. Statistical methodology for design, modeling and inference of such
experiments have been developed. In this chapter, we review some of those key developments, and
propose active learning approaches for modeling, optimization, contour estimation of computer
experiments with both types of inputs. Numerical studies are conducted to evaluate the performance of
the proposed methods in comparison with other existing methods.
\end{quote}

\begin{quote}
\end{quote}
 
\section{Introduction}\label{Section1}

Computer experiments are widely used to analyze computer simulators, which are mathematical models implemented in code to study complex systems \citep{sacks1989design}. These experiments provide a cost-effective and practical alternative to physical experiments.  They have been successfully applied across various scientific and engineering domains, including rocket booster simulations \citep{gramacy2020surrogates} and high-performance computing systems analysis \citep{cameron2019moana}.
One challenge in using computer experiments is that they can be computationally expensive in certain applications, with some simulations requiring hours or even days to generate a single output. To address this challenge, statistical emulators, such as Gaussian process (GP) models, are employed to approximate the behavior of computer simulators. These emulators significantly support the follow-up analysis and inference goals such as optimization \citep{jones1998efficient}, contour estimation \citep{ranjan2008sequential}, and prediction \citep{yang2020global}, among others.

Computer experiments often involve diverse data inputs, including scenarios where both quantitative and qualitative inputs are involved. As emulators, such as GP models, typically assume quantitative inputs, there is a growing need for advanced modeling approaches to accommodate mixed inputs. Recent developments have addressed this challenge following early work such as \cite{qian2008gaussian, han2009prediction,zhou2011simple}. For example, \cite{deng2017additive} introduced additive Gaussian process (AGP) models, incorporating an additive structure for qualitative inputs while assuming a multiplicative correlation between qualitative and quantitative inputs. Building on this foundation, \cite{xiao2021ezgp} proposed the easy-to-interpret Gaussian process (EzGP) model, which numerically outperforms AGP. Their approach eliminates the need to explicitly construct a correlation function for qualitative inputs, instead accounting for the effects of different level combinations through an additive GP structure. \cite{zhang2021mixed} focuses on the case when there is a large number of qualitative inputs. They proposed a sparse covariance estimation approach to estimate the covariance matrix for the mixed-input Gaussian process emulator. 
In addition, \cite{zhang2020latent} developed the latent variable Gaussian process (LvGP), a novel method that maps qualitative inputs to underlying numerical latent variables to define the correlation function. More recently, \cite{lin2024category} advanced GP models for mixed inputs by introducing a tree structure to partition the levels of qualitative inputs, resulting in enhanced model performance. These innovations have significantly expanded the capabilities of GP models to handle mixed input scenarios in computer experiments.

Another fundamental issue in computer experiments is determining how to select inputs for running computer codes. This is known as {\em experimental design}. Significant progress has been made in developing experimental designs for computer experiments involving mixed inputs. Notable examples include sliced Latin hypercube designs \citep{qian2012sliced, hwang2016sliced, yang2013construction} and marginally and doubly coupled designs \citep{deng2015design, he2017construction, he2017marginally, yang2023doubly}. These designs are space-filling and do not rely on specific models.
However, such model-independent designs can be inefficient because they do not explicitly take into account the study's objectives. To address this limitation, adaptive designs provide a more effective alternative. Adaptive designs select inputs based on the specific analysis goals, thereby reducing unnecessary computational time and costs. They typically begin with an initial design, fit a model to the available data, and iteratively update the model after each selection until the desired objective is achieved.

A critical step in adaptive designs is determining next input, which is selected based on a specific criterion aligned with the study’s objectives. Here we focus on commonly used adaptive design criteria for optimization, contour estimation, and prediction, although there are other objectives such as the estimation of a probability of failure \citep{bect2012sequential}. 
For optimization, widely discussed criteria include expected improvement (EI) \citep{jones1998efficient}, upper confidence bound (UCB) \citep{srinivas2009gaussian}, quantile-based \citep{picheny2013quantile} and entropy-based approaches \citep{hernandez2014predictive}. For contour estimation, \cite{ranjan2008sequential} extended the EI criterion originally proposed for optimization to contour estimation, while \cite{cole2023entropy} applied entropy-based designs, introducing a failure region to avoid complex integral evaluations. For prediction, \cite{yang2020global} use the EI criterion for multiple contours estimation for adaptive designs for prediction. In addition, \cite{mohammadi2022cross} proposed an adaptive design approach using leave-one-out cross-validation to enhance Gaussian process (GP) models for prediction.
Most of these criteria have been applied primarily to models with quantitative inputs. Notable exceptions include \cite{cai2024adaptive}, \cite{zhang2020bayesian}, and \cite{luo2024hybrid}, who considered adaptive designs for optimization with mixed inputs. Furthermore, \cite{Shahrokhian2024adaptive} proposed a comprehensive framework for adaptive designs for contour estimation in computer experiments involving mixed inputs.

The goal of this chapter is to survey recent developments in adaptive designs for contour estimation and optimization in computer experiments with mixed inputs and to provide a detailed comparison of these methods. In addition, we evaluate the performance of adaptive design criteria, originally proposed for prediction in computer experiments with only quantitative inputs \citep{yang2020global}, when applied to computer experiments with mixed inputs.

The rest of the chapter is organized as follows. Section \ref{Section2} provides the necessary background on GP models for computer experiments with both types of inputs, and reviews different adaptive designs used in models with quantitative inputs for optimization, contour estimation and prediction.  Section \ref{Section3} presents different adaptive designs for optimization and contour estimation for computer experiments with mixed inputs. We illustrate the comparison of these methods through the simulation studies in Section \ref{Section4}. Section \ref{Section5} concludes the paper.

\section{Background}\label{Section2}

This section provides the foundational background for adaptive designs in computer experiments involving both quantitative and qualitative inputs, with applications in contour estimation, optimization, and prediction. Section~\ref{Section2.1} reviews the easy-to-interpret Gaussian process (EzGP) model, a framework for emulating computer simulators with mixed inputs. Section~\ref{Section2.2} summarizes adaptive design strategies tailored to optimization, contour estimation, and prediction in experiments with quantitative inputs.

\subsection{Modeling}\label{Section2.1}
To review the EzGP, we first review Gaussian process models.
Consider a computer experiment with $p$ quantitative input variables. Suppose $i=1,\ldots,n$, design points $\bm{x}_i=(x_{i1},\ldots,x_{ip})^\top$ are used and the corresponding responses $y_1,\ldots,y_n$ are collected where $y_i=y(\bm{x}_i)$ and $\bm{x}_i\in (0,1)^p$. A Gaussian process model assumes that,  
\begin{align*}
    Y(\bm{x}_i)=F\bm{\beta}+Z(\bm{x}_i)~~~i=1,\ldots,n,
\end{align*}
where $F$ is a vector of the known basis function of the input, $\bm{\beta}$ is the vector of the regression coefficients, $Z(\bm{x}_i)$ is a GP with $E[Z(\bm{x}_i)]=0$,  $\mbox{Var}(Z(\bm{x}_i))=\sigma^{2}$ and $\mbox{Cov}(Z(\bm{x}_i),Z\bm{(x}_j))$\\$=\sigma^2 R_{ij}$. When the mean part $F\bm{\beta}$ equals to a constant $\mu$, it is known as ordinary kriging. There are many choices for the correlation functions. One of the most commonly used  correlation functions is the Gaussian correlation that is,
\begin{align}\label{Rij}
    R_{ij}=\mbox{Corr}(Z(\bm{x}_i),Z(\bm{x}_j))=\prod_{k=1}^{p}\exp\{-\theta_k(x_{ik}-x_{jk})^{2}\},
\end{align}
where the parameters $\bm{\theta}=(\theta_1,\ldots,\theta_p)^\top$ are the correlation parameters. 
A GP assumes $(Y_1,\ldots,Y_n)^\top$ follows a multivariate normal (MVN) distribution $N_n(\mu \bm{\mathds{1}}_n,\bm{\Sigma})$ where $\bm{\mathds{1}}_n$ is an $n$-vector of ones and $\bm{\Sigma}=\sigma^{2} R_{ij}$.
 To estimate the parameters, we can use the maximum likelihood approach. In this model we have $p+2$ parameters, $\mu$, $\sigma^2$ and $\bm{\theta}=(\theta_1,\ldots,\theta_p)^\top$. Let $\bm{y}=(y_1,\ldots,y_n)^\top$ denote the vector of $n$ responses, $\bm{R}$ denote the $n\times n$ matrix whose $(i,j)$ entry is $\mbox{Corr}(Y(x_i),Y(x_j))$. Then the likelihood function is, 
\begin{align}\label{GASPlike}
    L(\bm{\theta},\mu,\sigma^2)=\frac{1}{(2\pi)^{(n/2)}|\bm{\Sigma}|^{1/2}}\exp\{-\frac{1}{2\sigma^2}(\bm{y}-\bm{\mathds{1}}_n\mu)^\top \bm{R}^{-1}(\bm{y}-\bm{\mathds{1}}_n\mu)\}.
\end{align}
Note that the dependence of $\bm{\theta}$ is through the correlation function in (\ref{Rij}).
Given the correlation parameters $\bm{\theta}=(\theta_1,\ldots,\theta_p)^\top$, the maximum likelihood estimates of $\mu$ and $\sigma^2$ are given by,
\begin{align}\label{GASPestimates}
    \hat{\mu}=(\bm{\mathds{1}}_n^{\top}\bm{R}^{-1}\bm{\mathds{1}}_n)^{-1}(\bm{\mathds{1}}_n^\top\bm{R}^{-1}\bm{y})~~\mbox{and}~~\hat{\sigma}^2=\frac{(\bm{y}-\bm{\mathds{1}}_n\hat{\mu})^\top\bm{R}^{-1}(\bm{y}-\bm{\mathds{1}}_n\hat{\mu})}{n}.
\end{align}
By substituting the estimates in (\ref{GASPestimates}) into the likelihood function (\ref{GASPlike}), we get the likelihood function that only depends on $\bm{\theta}$ and can be maximized for $\bm{\theta}$. This model can now be used to make a prediction at any unsampled point $\bm{x}^*$. The predictive mean and the predictive variance are given by, 
\begin{align}\label{quant_predict}
\begin{split}
    \hat{\mu}(\bm{x}^*)&=\hat{\mu}+\bm{r}^\top\bm{R}^{-1}(\bm{y}-\bm{\mathds{1}}_n\hat{\mu}),\\
\hat{\sigma}^2(\bm{x}^*)&=\sigma^2(1-\bm{r}^\top\bm{R}^{-1}\bm{r}+\frac{(1-\mathds{1}_n^\top\bm{R}^{-1}\bm{r})^2}{\bm{\mathds{1}}_n^\top\bm{R}^{-1}\bm{\mathds{1}}_n}),
\end{split}
\end{align}
where $\bm{r}=(r_{1}(\bm{x}^*),\ldots,r_{n}(\bm{x}^*))^\top$, and $r_{i}(\bm{x}^*)=\mbox{Corr}(Y(\bm{x}^*),Y(\bm{x}_i))$.

For computer experiments with quantitative and qualitative inputs, \cite{xiao2021ezgp} proposed the EzGP model. Consider an $n$-run computer experiment with $p$ quantitative variables and $q$ qualitative factors. Let $x^{(k)}$ be the $k$th quantitative variable and $z^{(h)}$ be the $h$th qualitative factor having $m_h$ levels,  for $k=1,\ldots, p$, $h=1,\ldots, q$.  For $i=1,\ldots, n$, denote $\bm{x}_i = (x_{i1},\ldots, x_{ip})^\top$ and 
$\bm{z}_i = (z_{i1},\ldots, z_{iq})^\top$, and let $\bm{w}_i = (\bm{x}_i^\top,\bm{z}_i^\top)^\top$ be the $i$th input and $y_i$ be the corresponding output.  
To model the relationship between output $Y$ and an input $\bm{w} = (\bm{x}^\top,\bm{z}^\top)^\top$,  \cite{xiao2021ezgp} proposed the model, 
\begin{align}\label{ADGP}
    Y(\bm{x},\bm{z})=\mu+G_0(\bm{x})+G_{{z}^{(1)}}(\bm{x})+\ldots+G_{{z}^{(q)}}(\bm{x}),
\end{align}
where $\mu$ is a constant mean, $G_0$ and $G_{{z}^{(h)}} $'s are independent GPs with mean zero and covariance function $\bm{\phi}_0$ and $\bm{\phi}_h$, $h=1,\ldots,q$.  \cite{xiao2021ezgp} assumes that  $G_0$ is a standard GP with only quantitative inputs $\bm{x}$ with $\bm{\phi}_0$ given by, 
 \begin{align}\label{EzGpcov1}
     \bm{\phi}_0(\bm{x}_i,\bm{x}_j|\bm{\theta}_0)=\sigma_0^2\exp\{-\sum_{k=1}^{p}\theta_k^{(0)}(x_{ik}-x_{jk})^2\},
 \end{align}
\noindent where the correlation parameter $\bm{\theta_0}=(\theta_1^{(0)},\ldots,\theta_p^{(0)})^\top$ are all positive. They proposed the covariance function of $G_{{z}^{(h)}}$ such that, 
 \begin{align}\label{EzGpcov2}
 \begin{split}
     \bm{\phi}_h((\bm{x}_i^\top,z_{ih})^\top,(\bm{x}_j^\top,z_{jh})^\top|\bm{\Theta}^{(h)})&=\sigma_h^2\exp\{-\sum_{k=1}^{p}\theta^{(h)}_{kl_h}(x_{ik}-x_{jk})^2\}\bm{\mathds{1}}(z_{ih}=z_{jh}\equiv l_h),\\
 \end{split}
 \end{align}
where $z_{ih}$ and $z_{jh}$ are the $i$th and $j$th value of the input variable ${z}^{(h)}$,  $l_h=1,\ldots,m_h$, $m_h$ is the number of the levels of the qualitative input variable ${z}^{(h)}$, $\bm{\Theta}^{(h)}=(\theta^{(h)}_{kl_h})_{p\times m_h}$ is the matrix for the correlation parameters, and the indicator function is 1 when $z_{ih}=z_{jh}\equiv l_h$ and 0 otherwise. Therefore, from (\ref{EzGpcov1}) and (\ref{EzGpcov2}), the covariance function in the model in (\ref{ADGP}) is given by, 
\begin{align}\label{EZGPcov}
\begin{split}
 \bm{\phi}(\bm{w}_i,\bm{w}_j)&=\hbox{Cov}(Y(\bm{w}_i),Y(\bm{w}_j))\\
    &=\bm{\phi}_0(\bm{x}_i,\bm{x}_j|\bm{\theta}_0)+\sum_{h=1}^{q}\bm{\phi}_h(\bm{x}_i,\bm{x}_j|\bm{\Theta}^{(h)})\\
    &=\sigma_0^2\exp\{-\sum_{k=1}^{p}\theta_k^{(0)}(x_{ik}-x_{jk})^2\}\\
    &\ \ \ \ +\sum_{h=1}^{q}\sum_{l_h=1}^{m_h}\bm{\mathds{1}}(z_{ih}=z_{jh}\equiv l_h)\sigma_h^2\exp\{-\sum_{k=1}^{p}\theta^{(h)}_{kl_h}(x_{ik}-x_{jk})^2\}.
    \end{split}
\end{align}
Given the covariance function, there are  $2+p+q+p\sum_{h=1}^{q}m_h$ parameters $\mu$, $\sigma_0^2$, $\theta_k^{(0)}$, $\sigma_h^2$, and $\theta_{kl_h}^{(h)}$ for  $h=1,\ldots,q,~k=1,\ldots,p$ and $l_h=1,\ldots,m_h$. They can be estimated using the { maximum likelihood estimation} (MLE) method. After dropping the constant terms of the log-likelihood function, under the GP model in (\ref{ADGP}), the MLE method aims to minimize,
\begin{align}
   \log|\bm{\Phi}|+(\bm{y}-\mu \bm{\mathds{1}}_n)^{\top}\bm{\Phi}^{-1}(\bm{y}-\mu \bm{\mathds{1}}_n),
\end{align} 
with $\bm{\mathds{1}}_n$ being a column of $n$ ones, $\bm{\sigma}^2=(\sigma_0^2,\ldots,\sigma^2_q)^\top$ and $\bm{\Theta}=(\bm{\theta}^{(0)},\bm{\Theta}^{(1)},\ldots,\bm{\Theta}^{(q)})$ where $\bm{\theta}^{(0)}=(\theta_k^{(0)})_{ p \times 1}$, $\bm{\Theta}^{(h)}=(\theta_{kl_h}^{(h)})_{p\times m_h}$, $\bm{y}=(y_1,\ldots,y_n)^\top$ and $\bm{\Phi}=(\bm{\phi}(\bm{w}_i,\bm{w}_j))_{n\times n}$ with the covariance given in (\ref{EZGPcov}). For a given $\bm{\sigma}^2$ and $\bm{\Theta}$, the maximum likelihood estimate of $\mu$ is, 
\begin{align}
    \hat{\mu}=(\bm{\mathds{1}}_n^\top\bm{\Phi}^{-1}\bm{\mathds{1}}_n)^{-1}\bm{\mathds{1}}_n^\top\bm{\Phi}^{-1}\bm{y},
\end{align}
and $\bm{\sigma}^2$ and $\bm{\Theta}$ can be obtained from,
\begin{align*}
    \{\bm{\sigma}^2,\bm{\Theta}\}=\operatorname*{argmin}_{\bm{\sigma}^2,\bm{\Theta}}\{\log|\bm{\Phi}|+\bm{y}^\top\bm{\Phi}^{-1}\bm{y}-(\bm{\mathds{1}}_n^\top\bm{\Phi}^{-1}\bm{\mathds{1}}_n)^{-1}(\bm{\mathds{1}}_n\bm{\Phi}^{-1}\bm{y})^2\},
\end{align*}
using a minimization algorithm, such as \texttt{rgenoud} in \texttt{R} \citep{r-genoud}. Denote $Y_{*}=Y(\bm{w}^*)$ as the prediction of $Y$ at a new input $\bm{w}^*$. Given $\hat{\mu}$, $\bm{\sigma}^2$ and $\bm{\Theta}$, the predictive mean and the predictive variance at a new location $\bm{w}^{*}$ are given by, 
\begin{align}\label{ARSDC_dist}
\begin{split}
 E(Y_{*}|\bm{y})&=\mu_{*}=\hat{\mu}+\bm{r}_0^\top\bm{\Phi}^{-1}(\bm{y}-\hat{\mu}\bm{\mathds{1}}_n),\\
  \mbox{Var}(Y_{*}|\bm{y})&=\sigma^{{2}}_{*}=\sum_{i=0}^{q}\sigma^2_{i}-\bm{r}_0^\top\bm{\Phi}^{-1}\bm{r}_0+\frac{(1-\bm{\mathds{1}}_n^\top\bm{\Phi}^{-1}\bm{r}_0)^2}{\bm{\mathds{1}}_n^\top\bm{\Phi}^{-1}\bm{\mathds{1}}_n},
    \end{split}
\end{align}
where $\bm{r}_0$ is the covariance vector of $\bm{\phi}(\bm{w}^{*},\bm{w}_i)_{n\times 1}$ for $i=1,\ldots,n$.

\subsection{Adaptive Designs in Computer Experiments with Quantitative Inputs}\label{Section2.2}

This section examines the criteria for adaptive designs in computer experiments with quantitative inputs, focusing on their applications in optimization, contour estimation, and prediction. Before delving into the specifics of these criteria, we first outline the general steps of adaptive design, as summarized in Algorithm \ref{alg:general}.

Adaptive designs start with an initial design with $n_0$ runs. Then based on a fitted model we search for the next input by maximizing a particular criterion $C(\bm{x})$ depending on the goal of the study that will be discussed in this section. The stopping criterion can be the budget sample size or cost, or when there is little improvement in the prediction accuracy. 

\begin{algorithm}
\caption{Adaptive Designs/ Active Learning}\label{alg:general}
\begin{algorithmic}
\Require {a function $f(\cdot)$ providing output $y= f(\bm{x})$ for input $\bm{x}$, the initial design of size $n_0$, the budget sample size $N$, and a criterion $C(\bm{x})$ to search for the next point.}
\State  {Run a small initial design $\bm{X}_{n_0}=(\bm{x}^\top_1,\ldots, \bm{x}^\top_{n_0})$. Obtain $\bm{D}_{n_0}=(\bm{X}_{n_0},\bm{Y}_{n_0})$ where $\bm{Y}_{n_0}=(f(\bm{x}_{1}),\ldots,f(\bm{x}_{n_0}))$. Let $n=n_0$.}
\While{$n < N$}
\State {Obtain the surrogate model using $\bm{D}_n$.}
\State {Solve the criterion $C(\bm{x})$ based on the fitted model for the choice of $\bm{x}_{n+1}|\bm{D}_n:\bm{x}_{n+1}=\operatorname*{argmax}_{\bm{x} \in \chi} C(\bm{x})|\bm{D}_n$.}
    \State {Observe the response at the chosen location by $y_{n+1}=f(\bm{x}_{n+1})$.}
    \State  {Update $\bm{D}_{n+1}=\bm{D}_n\bigcup (\bm{x}_{n+1},y_{n+1})$, and set $n\leftarrow n+1$.}
\EndWhile\\
\textbf{Return} {the chosen design and the fitted model based on $\bm{D}_n$.}
\end{algorithmic}
\end{algorithm}

Note that Algorithm \ref{alg:general} is general in the sense that it works for models with only quantitative inputs, models with both types of inputs, and for the different goals of analysis including optimization, contour estimation and prediction and so on.

\subsubsection{Optimization}\label{Section2.2.1}

The optimization problem in adaptive designs involves identifying the input values that maximize or minimize the computer simulator. Commonly used criteria in optimization include the expected improvement (EI) criterion proposed by \cite{jones1998efficient} and the upper confidence bound (UCB) or lower confidence bound (LCB) criteria introduced by \cite{srinivas2009gaussian}.

The EI criterion, originally introduced by \cite{jones1998efficient}, is designed for minimizing black-box functions. The criterion measures the potential improvement of a solution over the current best result by accounting for both the predicted mean and the uncertainty provided by a surrogate model, such as a GP. Let  $f^{n}_{\min}=\min\{y_1,\ldots,y_n\}$, then improvement over $f^{n}_{\min}$ at an input location $\bm{x}$ can be defined as, 
\begin{align}\label{Io}
    I_o(\bm{x})=\max\{0,f^{n}_{\min}-y(\bm{x})\},
\end{align}
where $y(\bm{x})\sim N(\hat{\mu}(\bm{x}),\hat{\sigma}^2(\bm{x}))$. Note that the improvement function in \cite{jones1998efficient} is denoted by $I(\bm{x})$ but we use $I_o(\bm{x})$ instead to distinguish the improvement function  in later sections. The expected value of the improvement function in (\ref{Io}) is,
\begin{align}\label{EImax}
    E[I_o(\bm{x})]=(f_{\min}^n-\hat{\mu}(\bm{x}))\Phi(\frac{f_{\min}^{n}-\hat{\mu}(\bm{x})}{\hat{\sigma}(\bm{x})})+\hat{\sigma}(\bm{x})\phi(\frac{f_{\min}^{n}-\hat{\mu}(\bm{x})}{\hat{\sigma}(\bm{x})}),
\end{align}
where $\hat{\mu}(\bm{x})$ and $\hat{\sigma}(\bm{x})$ are the predictive mean, the predictive standard deviation given in (\ref{quant_predict}),  $\Phi$ and $\phi$ are the cumulative distribution function (CDF) and the probability density function (PDF) respectively. Let $\chi$ denote the design space throughout, the goal is to maximize (\ref{EImax}) such that,
\begin{align}\label{EIjones}
    \bm{x}_{n+1}=\operatorname*{argmax}_{\bm{x} \in \chi}{E[I_o(\bm{x})]}.
\end{align}

On the other hand, to balance the exploration and exploitation, \cite{srinivas2009gaussian} discussed two criteria, one for maximization where the UCB criterion selects the next candidate point to be, 
\begin{align*}
    \bm{x}_{n+1}=\operatorname*{argmax}_{\bm{x} \in \chi}\{\hat{\mu}(\bm{x})+{\rho}\hat{\sigma}(\bm{x})\},
\end{align*}
and the other for minimization that the LCB criterion selects the next point to be, 
\begin{align}\label{LCB}
\bm{x}_{n+1}=\operatorname*{argmin}_{\bm{x} \in \chi}\{\hat{\mu}(\bm{x})-{\rho}\hat{\sigma}(\bm{x})\}.
\end{align}
where $\rho \geq 0$ is tuning parameter, $\hat{\mu}(\bm{x})$ and $\hat{\sigma}(\bm{x})$ are the predictive mean, the predictive standard deviation given in (\ref{quant_predict}).

\subsubsection{Contour Estimation}\label{Section2.2.2}

The contour estimation problem involves identifying the inputs that correspond to a specified contour level of interest. Suppose $a$ is the contour level of interest and the contour is defined to be $S(a)=\{\bm{x}\in \chi:y(\bm{x})=a\}$. For computer experiments with quantitative inputs,  \cite{ranjan2008sequential} introduced the EI criterion with the improvement function defined as, 
\begin{align}\label{EIran}
    I(\bm{x})=\epsilon^2-\min\{(y(\bm{x})-a)^2,\epsilon^2\},
\end{align}
where $\epsilon=\alpha \hat{\sigma}(\bm{x})$ for $\alpha>0$, and $y(\bm{x})\sim N(\hat{\mu}(\bm{x}),\hat{\sigma}^2(\bm{x}))$ with the predictive mean $\hat{\mu}(\bm{x})$ and the predictive standard deviation $\hat{\sigma}(\bm{x})$ given in (\ref{quant_predict}). Let ${u}_1=(a-\hat{\mu}(\bm{x})-\epsilon)/\hat{\sigma}(\bm{x})$ and ${u}_2=(a-\hat{\mu}(\bm{x})+\epsilon)/\hat{\sigma}(\bm{x})$. Then  expected value of the improvement function in (\ref{EIran}) is,
\begin{align}\label{ranjan}
\begin{split}
    E[I(\bm{x})]&=[\epsilon^2-(\hat{\mu}(\bm{x})-a)^2-\hat{\sigma}^2(\bm{x})](\Phi({u}_2)-\Phi({u}_1))+\\&\hat{\sigma}^2(\bm{x})(u_2\phi({u}_2)-u_1\phi({u}_1))+2(\hat{\mu}(\bm{x})-a)\hat{\sigma}(\bm{x})(\phi({u}_2)-\phi({u}_1)),
\end{split}
\end{align}
where  $\Phi$ and $\phi$ are the CDF and the PDF of the standard normal distribution, respectively. The EI criterion in (\ref{ranjan}) chooses the next input, 
\begin{align*}
    \bm{x}_{n+1}=\operatorname*{argmax}_{\bm{x}\in \chi}{E[I(\bm{x})]}.
\end{align*}

 \cite{cole2023entropy} studied an entropy-based adaptive design for contour estimation. In their study, the problem depends on the definition of the failure region  which is characterized as an output $Y$ exceeding a particular contour level $a$. That is, for a limit state function $g(Y)=Y-a$, such that, 

\begin{align}\label{ECLdef}
    \mathbb{G}=\{\bm{x}\in \chi: g(Y(\bm{x}))>0\}~~\mbox{implying contour}~~ \mathbb{C}=\{\bm{x}\in\chi:g(Y(\bm{x}))=0\},
\end{align} 
where $\mathbb{G}$ is the failure region and $\mathbb{C}$ is the contour of interest. They introduced the entropy-based contour locator (ECL) criterion, 
\begin{align}\label{ELC}
\begin{split}
\medmath{\mbox{ECL}(\bm{x}|g)}=&\medmath{-P(g(Y(\bm{x}))>0)\log P(g(Y(\bm{x}))>0)-P(g(Y(\bm{x}))\leq 0)\log P(g(Y(\bm{x}))\leq 0)}\\
    &\medmath{=-(1-\Phi(\frac{\hat{\mu}(\bm{x})-a}{\hat{\sigma}(\bm{x})}))\log(1-\Phi(\frac{\hat{\mu}(\bm{x})-a}{\hat{\sigma}(\bm{x})}))-\Phi(\frac{\hat{\mu}(\bm{x})-a}{\hat{\sigma}(\bm{x})})\log(\Phi(\frac{\hat{\mu}(\bm{x})-a}{\hat{\sigma}(\bm{x})}))},
\end{split}
\end{align}
where $\Phi$ is the CDF of the standard normal distribution, for any limit state function $g(Y)=b(Y-a)$ where $b\in\{-1,1\}$.  It shall be noted because of symmetry in (\ref{ELC}), the criterion works identically for $g(Y)=b(Y-a)$ where $b\in\{-1,1\}$, meaning that the criterion is unchanged for $b=-1$ or $b=1$.  
The ECL criterion chooses the next input as,
\begin{align*}
    \bm{x}_{n+1}=\operatorname*{argmax}_{\bm{x}\in \chi}\mbox{ECL}(\bm{x}|g).
\end{align*}

\subsubsection{Prediction}\label{Section2.2.3}

Prediction aims to accurately estimate the model output and ensure reliable prediction. While several adaptive design approaches exist for enhancing prediction, this section focuses on the methods employed in our numerical studies. We review the EI multiple contours (EI-MC) criterion and the EI single contour (EI-SC) criterion proposed by \cite{yang2020global}, which have been developed as adaptive design strategies for prediction.

\cite{yang2020global} consider the improvement function in (\ref{EIran}) for a given integer $c>0$ and a set of scalar $a_1,\ldots,a_{c}\in[y_{\min},y_{\max}]$. Suppose we are interested in estimating $S(a_1),\ldots,S(a_{c})$, where $a_1,\ldots,a_{c}\in[y_{\min},y_{\max}]$. The values of $y_{\min}$ and $y_{\max}$ are unknown contours in general but it can be estimated by a large input from the design domain. They suggest, $1000p$ design points. Note that $a_1<\ldots<a_{c}$ then the improvement function in (\ref{EIran}) is adjusted such that, 
\begin{align}\label{EIMC_im}
     I_c(\bm{x})=\epsilon^2-\min\{(y(\bm{x})-a_1)^2,\ldots,(y(\bm{x})-a_{c})^2,\epsilon^2\},
\end{align}
where $y(\bm{x})\sim N(\hat{\mu}(\bm{x}),\hat{\sigma}^2(\bm{x}))$, $\epsilon=\alpha\hat{\sigma}(\bm{x})$ for some positive constant $\alpha$. The improvement function will be nonzero if $(y(\bm{x})-a_j)^2<\epsilon^2$ for some $j$. Therefore the expectation of the improvement function in (\ref{EIMC_im}) is sum of the individual contour estimation  EI criterion in (\ref{ranjan}) over $c$ cases given by, 
\begin{align}\label{EIMC}
\begin{split}
    E[I_c(\bm{x})]&=\sum_{j=1}^{c}[\epsilon^2-(\hat{\mu}(\bm{x})-a_j)^2-\hat{\sigma}^2(\bm{x})](\Phi({u}_2)-\Phi({u}_1))+\\&\hat{\sigma}^2(\bm{x})(u_2\phi({u}_2)-u_1\phi({u}_1))+2(\hat{\mu}(\bm{x})-a_j)\hat{\sigma}(\bm{x})(\phi({u}_2)-\phi({u}_1)),
\end{split}
\end{align}
where  $\phi$ and $\Phi$ are the PDF and the CDF of the standard normal distribution, respectively. The EI-MC criterion chooses the next input, 
\begin{align*}
    \bm{x}_{n+1}=\operatorname*{argmax}_{\bm{x}\in \chi}{E[I_c(\bm{x})]}.
\end{align*}

 \cite{yang2020global} considered an other way for EI-SC, that they use the EI function in (\ref{EIran}). They consider an automatic approach to select $a_j$ in each adaptive design. The contour level is selected from the candidate points to select the next input. Let the candidate points for the next input be $\bm{x}^*_1,\ldots,\bm{x}^*_m$ and,
\begin{align*}
     \bm{x}^{*}_{opt}=\operatorname*{argmax}_{1<i<m}{\sigma^2(\bm{x}_i^*)}.
\end{align*}
They set $a_j=\hat{y}(\bm{x}^{*}_{opt})$ and use the EI criterion in (\ref{ranjan}) given $a_j$ in each adaptive design.

\section{Active Designs for Computer Experiments with Quantitative and Qualitative Inputs}\label{Section3}

This section examines adaptive designs for computer experiments involving both quantitative and qualitative inputs, focusing on the goals of optimization and contour estimation. Incorporating qualitative inputs introduces several unique challenges. First, while criteria in experiments with quantitative inputs often depend solely on the predictive mean and variance, the influence of qualitative inputs on these criteria becomes difficult to quantify. Second, optimizing the criteria is more complex because the search space encompasses both continuous and discrete regions. These challenges necessitate innovative methodologies for adaptive design in experiments with mixed inputs. The following subsections review recent advancements in adaptive designs for optimization and contour estimation in such contexts.

\subsection{Optimization}\label{Section3.1}

 \cite{cai2024adaptive} and \cite{luo2024hybrid}  have explored optimization methods for computer experiments with mixed inputs. In this subsection, we review their approaches and provide a numerical example to illustrate their methodologies.

\cite{cai2024adaptive} introduced the {\em adaptive-region sequential design} (ARSD) approach for addressing optimization problems in computer experiments with mixed inputs. The ARSD method selects the next candidate point to be 
\begin{align}\label{ARSD}
    \bm{w}_{n+1}=\operatorname*{argmin}_{\bm{w}\in \mathbb{A}_n}\{\hat{\mu}(\bm{w})-\rho\hat{\sigma}(\bm{w})\},
\end{align}
where $\mathbb{A}_n \subset \mathbb{A}$ is the adaptive design region as,
\begin{align}\label{Adaptive}
    \mathbb{A}_n=\{\bm{w}\in \mathbb{A}:\hat{\mu}(\bm{w})-\sqrt{\beta_{0|n}}\hat{\sigma}(\bm{w})\leq \min[\hat{\mu}(\bm{w})+\sqrt{\beta_{0|n}}\hat{\sigma}(\bm{w})]\},
\end{align}
$\rho \geq 0$ is a tuning parameter, $\mathbb{A}=\{(\bm{x},\bm{z})|\bm{x}\in \chi,\bm{z}\in \bm{Z}\}$ is the whole design space and $\bm{Z}$ is all level combinations of the qualitative inputs, $\hat{\mu}(\bm{w})$ and $\hat{\sigma}(\bm{w})$ are the predicted mean and the predicted standard deviation, $\beta_{0|n}=2\log(\frac{\pi^2n^2M}{6\alpha})$, $\alpha\in(0,1)$ and $M=\prod_{j=1}^{q}m_j$, and $m_j$ is the number of level of the $j$th qualitative input variable. 
The criterion in (\ref{ARSD}) is essentially a LCB criterion tailored for minimization problems, combined with an adaptive search region defined in (\ref{Adaptive}). This adaptive region includes inputs whose predictive lower bounds are smaller than the minimum predictive upper bound across the entire design space. By narrowing the search region, this approach significantly reduces computational time. Moreover, the efficiency loss is minimal, as \cite{cai2024adaptive} provides a theoretical guarantee that the solution to the minimization problem lies within the adaptive search region $\mathbb{A}_n$ with  a high probability.

\cite{luo2024hybrid} introduces a novel sequential approach for selecting inputs to optimize a black-box function with mixed variables. While not specifically designed for computer experiments with mixed inputs, their method can effectively address optimization problems in such settings. The approach is described as {\em Hybrid} because it combines Monte Carlo tree search (MCTS) \citep{kocsis2006bandit} for exploring qualitative inputs with GPs for modeling quantitative inputs.
A key contribution of their work is the introduction of a family of candidate kernels for mixed variables, along with a numerically stable and efficient dynamic kernel selection criterion. This criterion identifies the optimal mixed-variable GP surrogate at each stage, which is then used to determine the next set of quantitative inputs.

To outline the methodology, we begin with a discussion of MCTS. MCTS incrementally constructs a search tree through an iterative process that has four key steps: selection, expansion, simulation, and back-propagation.
\begin{itemize}
\item Selection: The algorithm navigates the tree using a policy that balances exploration and exploitation, often guided by criteria such as the Upper Confidence Bound for Trees (UCT) introduced by \cite{auer2002using}.
\item Expansion: New nodes representing previously unexplored states are added to the tree.
\item Simulation: The newly expanded nodes are evaluated using a reward function, defined in accordance with the specific objectives of the study.
\item Back-propagation: The results from the simulation are propagated back up the tree, updating value estimates for all nodes visited during the iteration.
\end{itemize}

In their study, MCTS is implemented as a one-layer tree with leaves corresponding to level combinations of qualitative inputs. A path along a tree with $L$
 layers is represented as a string $\mathbf{s} = (s_1, \ldots, s_L)$ of length $L$, where each  $\mathbf{s}$
  denotes a node on the path. During each search, the algorithm conditions on the path $(s_1, \ldots, s_i)$, $C$ and computes a reward function value. This information is back-propagated through the tree using update strategies to guide subsequent decisions in the adaptive design process.
The update strategy follows the UCT principle:
\begin{align}\label{UCB-MCTS}
    \bar{r}(s_1, \ldots, s_i) + C \sqrt{\frac{\log(n(s_1, \ldots, s_{i-1}))}{n(s_1, \ldots, s_i)}},
\end{align}
\noindent where $\bar{r}(s_1, \ldots, s_i)$ is the average reward from all evaluations along the path $(s_1, \ldots, s_i)$, 
$C$ is a pre-specified constant controlling the exploration-exploitation trade-off, and $n(\mathbf{s})$ represents the number of times the path 
$\mathbf{s}$  has been visited. New qualitative variable values are then determined by maximizing the  criterion in (\ref{UCB-MCTS}).  Once the new qualitative values have been chosen, they follow
an update strategy, by updating $n(\mathbf{s})$ and $\bar{r}(\mathbf{s})$ where $\mathbf{s}$ denotes a vector like $(s_1, \ldots, s_i)$, and the full path ($s_1, \ldots, s_L$) corresponds
to a qualitative variable value and propagates the
updated value along the complete path for each non-leaf node.
The updated value is propagated along the entire path for each non-leaf node. Once a new sample location is evaluated, the reward and visit count for all nodes along the tree path leading to the sample are updated. For each node $s_i$ in the
path, the visit count $n(s_1, \ldots, s_i)$  is increased by 1, and the average reward $\bar{r}(s_1, \ldots, s_i)$  is also updated to reflect the new sample’s reward.
 
After choosing the new qualitative variable values, they dynamically select the mixed-variable GP surrogate. 
At each adaptive design, a kernel $\bm{K}$ is selected from a pool of candidates, including additive and multiplicative kernels. The selection is guided by two metrics, one is the log marginal likelihood, which quantifies the kernel’s fit to the observed data, and the other is the acquisition function value, which represents the potential improvement. Each kernel is ranked by its log-likelihood \(R_P(\bm{K})\) and acquisition function \(R_A(\bm{K})\), and the combined rank is computed as,
\begin{equation}
    R(\bm{K}) = R_P(\bm{K}) + \alpha R_A(\bm{K}),
\end{equation}
where \(\alpha \in [0, 1]\) controls the balance between exploitation \(R_P(\bm{K})\) and exploration \(R_A(\bm{K})\). By default, \(\alpha = 0.5\), or \(\alpha\) can adapt over iterations as, $\alpha = \frac{2 \cdot i}{N}$, where \(i\) is the current iteration, and \(N\) is the total sampling budget. The kernel with the highest \(R(\bm{k})\) is selected and used to update the GP surrogate. Now given  the input and the selected surrogate, they use the EI criterion in (\ref{EIjones}) to identify the next quantitative variable values. For rest of the work we call this method as {\em Hybrid}.

To illustrate the methodology of each adaptive design for optimization, we consider a minimization problem in the following computer experiment with one qualitative input variable and one quantitative input variable. We examine the performance of adaptive designs approaches using the EI in (\ref{EIjones}), LCB in (\ref{LCB}), ARSD in (\ref{ARSD}) criteria and Hybrid method by \cite{luo2024hybrid} respectively.

\begin{example}\label{ex3.1}
Consider a computer experiment with $p=1$ quantitative input $x$ and $q=1$ qualitative input $z$ and the computer model is represented by, 
\begin{align}\label{Example1}
  f=\begin{cases}
      2-\cos(2\pi {x}), & \text{if}\ {z}=1\\  1-\cos(4\pi {x}), & \text{if}\ {z}=2\\ \cos(2\pi {x}), &  \text{if}\ {z}=3.
    \end{cases}
\end{align}
Here $-1$ is the minimum of the response surface and $3$ is the maximum of the response surface. The solution that minimizes the computer model is \(\bm{w} = (0.5, 3)\). 

Figure \ref{figure1} illustrates the adaptive designs using EI, ARSD, LCB, and Hybrid, along with the one-shot design, for $n_0 = 9$ and $N = 15$ in one simulation to find the minimum of the responses given by the computer model.  The dashed lines show the underlying true response of (\ref{Example1}) and the solid orange line indicates the minimum of the true responses is -1. The initial points are presented as circled dots in each level, corresponding to their colors, while the added points are labeled numerically. Figure \ref{figure1}(d) starts with a single initial point, with the remaining \(N = 15\) points adaptively added and labeled numerically using the Hybrid method. As shown in Figures \ref{figure1}(a)-(d), the added points in the adaptive designs are  mostly selected at level 3 and are located close to the target input \(\bm{w} = (0.5, 3)\). In contrast, as displayed in Figure \ref{figure1}(e), the minimum of the responses given by the one-shot design is given by \(\bm{w} = (0.25, 3)\) so that the corresponding minimum is close to 0. 

\begin{figure}[htbp]
\centering
\includegraphics[scale=0.54]{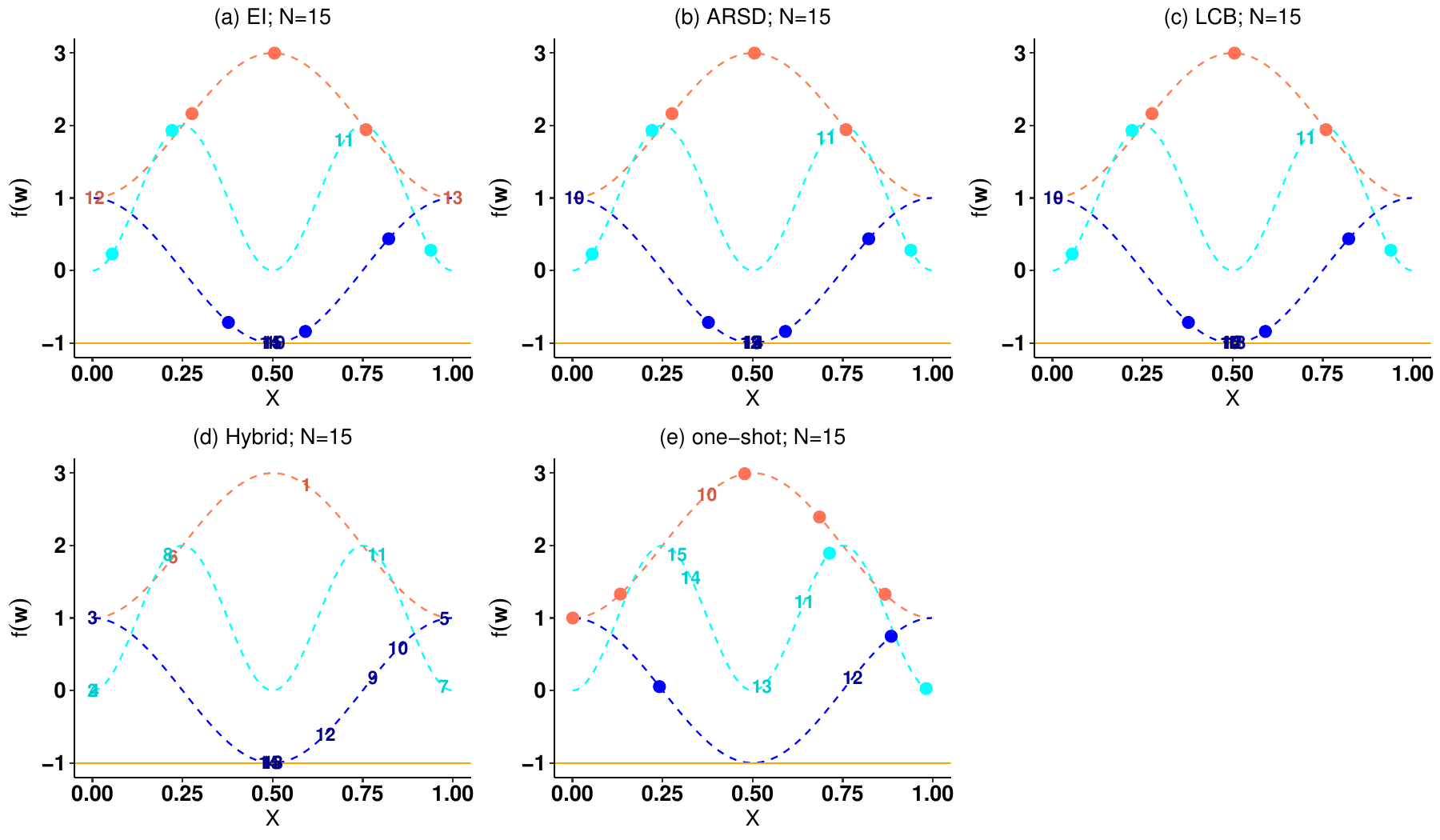}
\caption{\footnotesize Adaptive designs versus one-shot designs for Example 1 for (a) EI (b) LCB (c) ARSD (d) Hybrid (e) one-shot for $n_0=9;N=15$ for one simulation.}\label{figure1}
\end{figure}
\end{example}

\subsection{Contour Estimation}\label{Section3.2}

This subsection discusses adaptive design approaches for contour estimation in computer experiments involving both quantitative and qualitative input variables. To the best of our knowledge, \cite{Shahrokhian2024adaptive} is the only work that considers adaptive designs for the contour estimation in computer experiments with both quantitative and qualitative inputs. In this subsection, we review the method  proposed by \cite{Shahrokhian2024adaptive}. Let 
$a$ denote the contour level of interest; the objective is to identify the contour $\{\bm{w}: y(\bm{w}) = a\}$.  Let the lower bound (LB) and the upper bound (UB) be given by,
\begin{align}\label{LBUB}
\begin{split}
    &\mu_{h}^{L}(\bm{w})=|\hat{\mu}(\bm{w})-a| - \sqrt{\beta_{0|n}}\hat{\sigma}(\bm{w}),\\&\mu_{h}^{U}(\bm{w})=|\hat{\mu}(\bm{w})-a| + \sqrt{\beta_{0|n}}\hat{\sigma}(\bm{w}),
   \end{split}
    \end{align}
where $\hat{\mu}(\bm{w})$ is the predictive mean, $\hat{\sigma}(\bm{w})$ is the predictive standard deviation defined in (\ref{ARSDC_dist}),  $\beta_{0|n}=2\log(\frac{\pi^2n^2 M}{6\alpha})$, $\alpha\in(0,1)$, $M=\prod_{j=1}^{q}m_j$ and $m_j$ is the number of levels of the $j$th qualitative input variable. Note that, the choice of $\beta_{0|n}$ is from \cite{cai2024adaptive}. Let $\mathbb{A}$ be the whole design space, \cite{Shahrokhian2024adaptive} proposed to divide $\mathbb{A}$ into two disjoint groups, 
\begin{align}
    &\mathbb{A}_{1}=\{\bm{w}\in \mathbb{A}:|\hat{\mu}(\bm{w})-a|-\sqrt{\beta_{0|n}}\hat{\sigma}(\bm{w})>0\},\label{ACCEA}\\
    &\mathbb{A}_2=\{\bm{w}\in \mathbb{A}:|\hat{\mu}(\bm{w})-a|-\sqrt{\beta_{0|n}}\hat{\sigma}(\bm{w}) \leq 0\}.\label{A_3}
        \end{align}
Clearly,  $\mathbb{A}= \cup_{i=1}^2 \mathbb{A}_i$. For $\mathbb{A}_1$ in (\ref{ACCEA}), consider the candidate point given by, 
\begin{align}\label{criterion_min}
        \bm{w}_{n+1}=\operatorname*{argmax}_{\bm{w}\in\mathbb{A}_{1,\min}}\hat{\sigma}(\bm{w}),
\end{align}
where
\begin{align}\label{A_min}
\begin{split}
    \mathbb{A}_{1,\min}=\{\bm{w}\in \mathbb{A}_{1}:&|\hat{\mu}(\bm{w})-a|-\sqrt{\beta_{0|n}}\hat{\sigma}(\bm{w}) \leq\\&\min_{\bm{w}\in\mathbb{A}}[|\hat{\mu}(\bm{w})-a|+\sqrt{\beta_{0|n}}\hat{\sigma}(\bm{w})] \}.
\end{split}
 \end{align}
For $\mathbb{A}_2$ in (\ref{A_3}), let the candidate point be,
\begin{align}\label{criterion_A_3}
    \bm{w}_{n+1}=\operatorname*{argmax}_{\bm{w}\in \mathbb{A}_{2}} \mbox{ECL}[\bm{w}|g],
\end{align}
with the ECL criterion in (\ref{ELC}).  

Among the two candidate points selected from groups $\mathbb{A}_1$ and $\mathbb{A}_2$, \cite{Shahrokhian2024adaptive} select the next input with a larger value of $\frac{\hat{\sigma}(\bm{w})}{\max(\delta,|\hat{\mu}(\bm{w})-a|)}$, where $\delta$ is a small positive value. \cite{Shahrokhian2024adaptive} highlighted that the choice of $\delta$ plays a critical role in balancing exploration and exploitation. A very small $\delta$ strongly favors exploitation, selecting inputs primarily, if not exclusively, from  $\mathbb{A}_2$. Conversely, a very large $\delta$ shifts the focus entirely to exploration, effectively relying only on the predictive variance and resulting in excessive input selection from $\mathbb{A}_1$. To achieve a balance between exploration and exploitation, $\delta$ should be neither too large nor too small, with its optimal value depending on the range of responses. In the numerical examples, the recommended value suggested by \cite{Shahrokhian2024adaptive} will be used.
In general, it is desirable to select points whose predictive mean is closer to the contour level $a$ while having a larger predictive variance. However, this approach tends to empirically favor selection from a single group after several steps in the sequential procedure. To address this, the denominator 
${\max(\delta,|\hat{\mu}(\bm{w})-a|)}$ is introduced to encourage exploration of the design space. This adjustment ensures that when the predictive means of candidate points are close to the contour level, the point with the larger predictive variance is chosen.

 Algorithm \ref{alg:RCC} given in \cite{Shahrokhian2024adaptive} describes the adaptive design approach for finding follow-up points for contour estimation.  Their algorithm applies the idea of the adaptive search region in the ARSD approach in the design space $\mathbb{A}_{1}$, and the ECL criterion in the design space $\mathbb{A}_{2}$. 
 They named their approach  {\em region-based cooperative
criterion} (RCC).

\begin{algorithm}
\caption{Region-based Cooperative Contour Estimation}
\label{alg:RCC}
\begin{algorithmic}
\State{ Run a small initial design $\bm{W}_{n_0}=(\bm{w}^T_1,\ldots, \bm{w}^T_{n_0})$. Obtain $\bm{D}_{n_0}=(\bm{W}_{n_0},\bm{Y}_{n_0})$ where $\bm{Y}_{n_0}=(f(\bm{w}_{1}),\ldots,f(\bm{w}_{n_0}))$. Let $n=n_0$.}
\While{$n < N$}
\State{Fit the EzGP model using $\bm{D}_n$.}
\State{Divide the search space into the two groups $\mathbb{A}_1$ and $\mathbb{A}_2$.}
\State{For $\mathbb{A}_1$ defined in (\ref{ACCEA}), find the input that minimizes the criterion in (\ref{criterion_min}).}
\State{For $\mathbb{A}_2$ defined in (\ref{A_3}), find the input that maximizes the criterion in (\ref{criterion_A_3}).}
\State{Choose the input from the two chosen points that $\frac{\hat{\sigma}(\bm{w})}{\max(\delta,|\hat{\mu}(\bm{w})-a|)}$ is larger.}
\State{Run the experiment at the chosen location and let $y_{n+1}=f(\bm{w}_{n+1})$.}
\State{Update $\bm{D}_{n+1}=\bm{D}_n\bigcup (\bm{w}_{n+1},y_{n+1})$, and set $n\leftarrow n+1$.}
\EndWhile\\
\textbf{Return}  {Extract the estimated contour for $\hat{S}=\{\bm{w}\!:\hat{y}(\bm{w})=a\}$ from the estimated response surface.}
\end{algorithmic}
\end{algorithm}

Although the LCB or UCB criteria in (\ref{LCB}) and the ARSD criterion in (\ref{ARSD}) are not extended for contour estimation, they can be used. The LCB criterion for contour estimation (LCB-C) chooses the next input as,
\begin{align*}
        \bm{w}_{n+1}=\operatorname*{argmin}_{\bm{w}\in\mathbb{A}}\{|\hat{\mu}(\bm{w})-a|-\rho\hat{\sigma}(\bm{w})\},
\end{align*}
where $\rho$ is a tuning parameter and $\mathbb{A}$ is the whole design space. Note that the LCB criterion was proposed for the problem of global optimization, but we can apply it to the contour estimation problem by using $|\hat{\mu}(\bm{w})-a|$ rather than $\hat{\mu}$. 

Similarly, the ARSD approach in (\ref{ARSD}) for optimization can also be modified for contour estimation. That is, we can choose the next input to be, 
\begin{align*}
        \bm{w}_{n+1}=\operatorname*{argmin}_{\bm{w}\in\mathbb{A}^{*}}\{|\hat{\mu}(\bm{w})-a|-\rho\hat{\sigma}(\bm{w})\},
\end{align*}
where $\rho$ is a tuning parameter, and 
\begin{align*}
\mathbb{A}^{*}=\{\bm{w}\in \mathbb{A}:|\hat{\mu}(\bm{w})-a|-\sqrt{\beta_{0|n}}\hat{\sigma}(\bm{w}) \leq \min_{\bm{w}\in\mathbb{A}}[|\hat{\mu}(\bm{w})-a|+\sqrt{\beta_{0|n}}\hat{\sigma}(\bm{w})] \}.
 \end{align*}
We denote this approach as ARSD for contour estimation (ARSD-C).

\begin{figure}[htbp]
\centering
\includegraphics[scale=0.54]{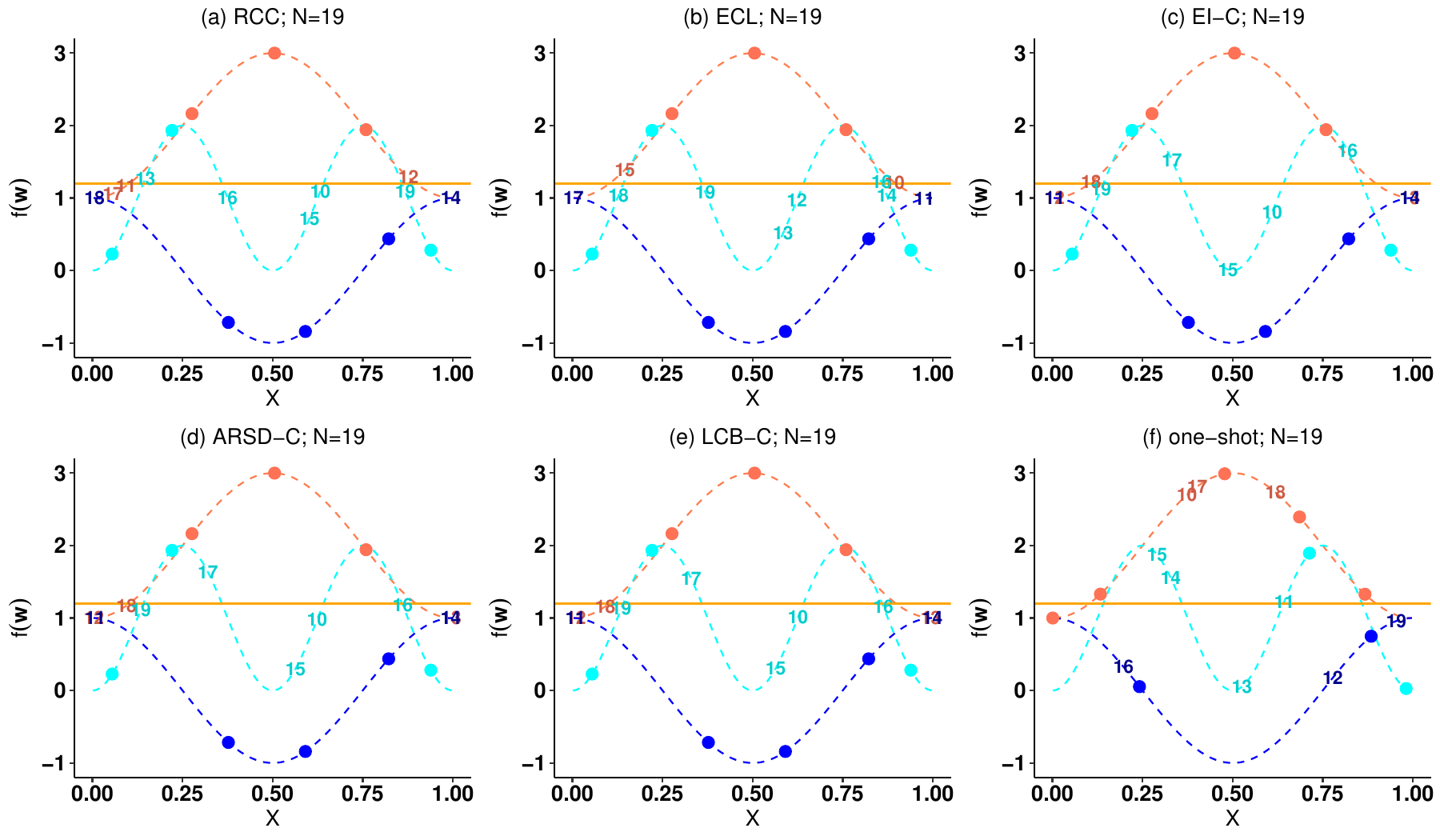}
\caption{\footnotesize Adaptive designs versus one-shot designs for Example 1 for (a) RCC, (b) ECL,  (c) EI-C, (d) ARSD-C (e) LCB-C, (f) one-shot  for $n_0=9; N=19$ for one simulation when $a=1.2$.}\label{figure2}
\end{figure}

\noindent {\em \textbf{Example 1 continued:}  Similar to the adaptive designs for optimization discussed in Section \ref{Section3.1}, this section illustrates various adaptive designs for contour estimation and compares them with one-shot designs. Figure \ref{figure2} presents the adaptive designs RCC, ECL, EI-C, ARSD-C, and LCB-C, alongside the one-shot designs, for $n_0=9$ and  $N=19$ in a simulation where 
$a=1.2$.  In the figure, the initial points are shown as circled dots, grouped according to their levels, while the follow-up points are labeled sequentially from 10 to 19. As seen in Figure \ref{figure2} (a)-(e), the adaptive designs primarily allocate additional points near the orange solid line, effectively targeting the contour corresponding to $a=1.2$. In contrast, the one-shot design in Figure \ref{figure2}~(f) distributes several points at level 3, many of which do not correspond to the desired contour level  $a=1.2$.
} 

\section{Numerical Studies}\label{Section4}

This section presents numerical examples to illustrate various adaptive designs for optimization, contour estimation, and prediction in computer experiments involving both quantitative and qualitative inputs.
For all adaptive designs, the process begins with an initial design with $n_0$ runs. The next input is selected by evaluating a large set of candidate points in  the design space. These candidate points include all possible combinations of the qualitative input levels, along with 100 random Latin hypercube designs (LHDs) \citep{mckay2000comparison} generated for each level combination in the design space.
In addition, one-shot designs are constructed by combining random LHDs for the quantitative variables with a (nearly) balanced sample across all level combinations of the qualitative factors. The random LHDs are generated using the \texttt{randomLHS} function from the \texttt{R} package \texttt{lhs} \citep{r-core}. To model the computer experiment with mixed inputs, we use the \texttt{EzGP} package in \texttt{R} \citep{r-EzGP}.

Section \ref{Section4.1} presents the simulation studies for contour estimation, while Section \ref{Section4.2} focuses on optimization, and Section \ref{Section4.3} explores the approaches for prediction. For each problem, we examine Example \ref{ex3.1} presented in Section 3, \nameref{ex3.2} and \nameref{ex3.3} as detailed below.   All examples and implementations may be found on the Github repository  \url{https://github.com/devonlin/Handbook-of-Statistics-Chapter}.

\subsection*{Example 2}\label{ex3.2}
We consider a computer model with $p=2$ quantitative inputs $\bm{x}=(x_1,x_2)$ and $q=2$ qualitative input $\bm{z}=(z_1,z_2)$. The computer model is adapted from \cite{xiao2021ezgp} with reduced dimensions and is  represented by, 
{\small
\begin{align*}
  i({z}_1)=
    \begin{cases}
      x_1+x_2^2, & \text{if}\ z_1=1 \\
      x_1^2+x_2, & \text{if}\ z_1=2\\
      x_1^2+x_2^2,&  \text{if}\ z_1=3,
    \end{cases}~~
  g({z}_2)=
    \begin{cases}
     \cos( x_1)+\cos(2 x_2), & \text{if}\ z_2=1 \\
      \cos( 2x_1)+\cos(x_2), & \text{if}\ z_2=2\\
     \cos(2x_1)+\cos(2x_2),&  \text{if}\ z_2=3,
    \end{cases}\\
    f=i(z_1)+g(z_2).
\end{align*}
}
\noindent Here $1$ is the minimum of the response surface and $2.7$ is the maximum of the response surface.

\subsection*{Example 3}\label{ex3.3}
We consider a computer model with $p=3$ quantitative inputs $\bm{x}=(x_1,x_2,x_3)$ and $q=3$ qualitative input $\bm{z}=(z_1,z_2,z_3)$ and the computer model \cite{xiao2021ezgp} is represented by, 
{\small
\begin{align*}
  i({z}_1)=
    \begin{cases}
      x_1+x_2^2+x_3, & \text{if}\ z_1=1 \\
      x_1^2+x_2+x_3, & \text{if}\ z_1=2\\
      x_3+x_1+x_2^2,&  \text{if}\ z_1=3,
    \end{cases}~~
  g({z}_2)=
    \begin{cases}
     \cos( x_1)+\cos(2 x_2)+\cos(x_3), & \text{if}\ z_2=1 \\
      \cos( x_1)+\cos(2 x_2)+\cos(x_3), & \text{if}\ z_2=2\\
     \cos(2x_1)+\cos(x_2)+\cos(x_3),&  \text{if}\ z_2=3,
    \end{cases}
\end{align*}
\begin{align*}
  h({z}_3)=
    \begin{cases}
       \sin(x_1)+\sin(2 x_2)+\sin(x_3), & \text{if}\ z_3=1 \\
      \sin(x_1)+\sin(2 x_2)+\sin(x_3), & \text{if}\ z_3=2\\
     \sin(2x_1)+\sin(x_2)+\sin(x_3),&  \text{if}\ z_3=3,
    \end{cases}~~
    f=i(z_1)+g(z_2)+h(z_3).
\end{align*}
}
\noindent Here $3$ is the minimum of the response surface and $6.7$ is the maximum of the response surface.

\subsection{Optimization}\label{Section4.1}

We evaluate the performance of adaptive designs for minimization using the EI, LCB, ARSD, and Hybrid methods discussed in Section \ref{Section2.1} and Section \ref{Section3.1}, alongside one-shot designs. To assess optimization accuracy, the best minimum value of the true surface obtained through the adaptive design is measured. For the ARSD method described in (\ref{ARSD}), the tuning parameter $\rho = 2$ is set as specified to ensure effective performance.

\begin{figure}[H]
\centering
\includegraphics[scale=0.6]{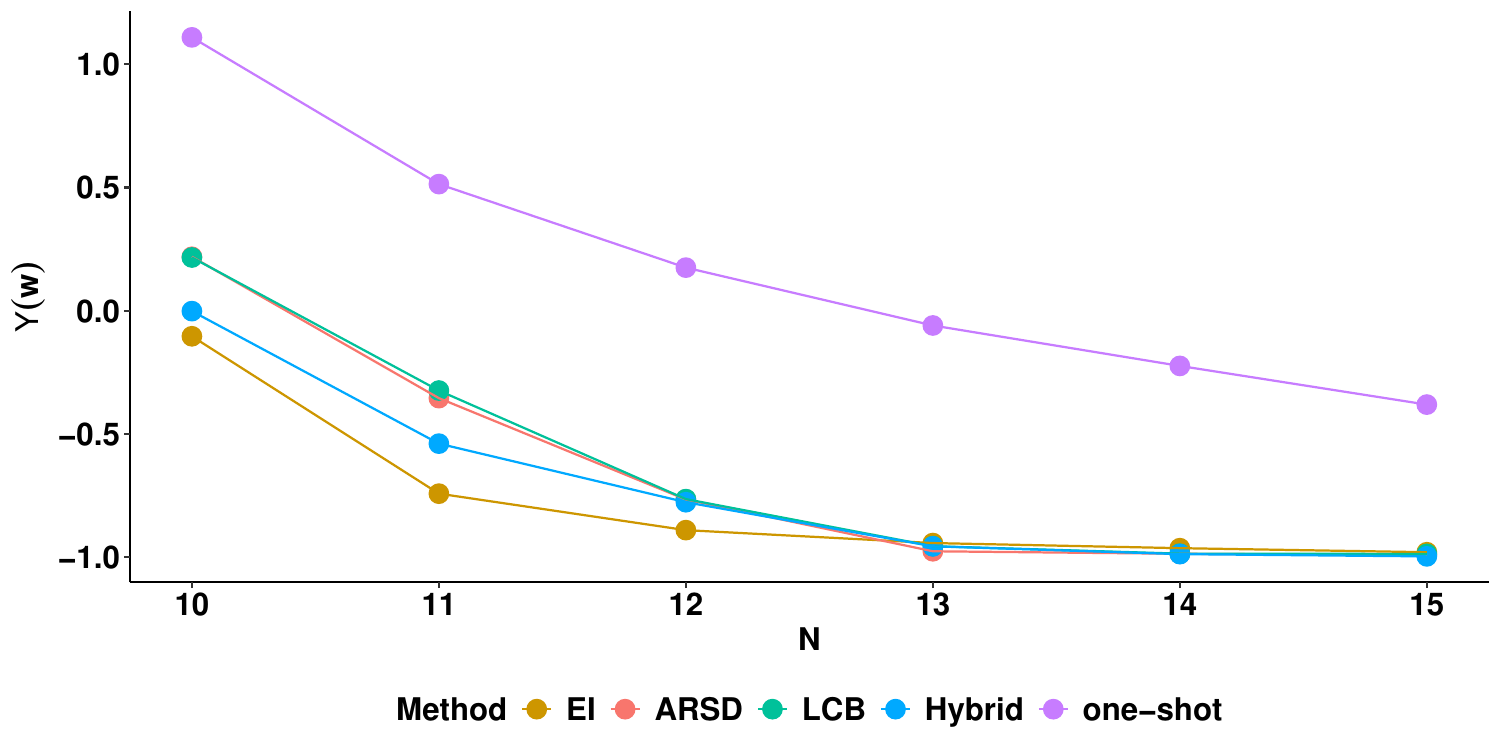}
\caption{\footnotesize The average of the best obtained minimum values of the true response surface in Example \ref{ex3.1} for the adaptive designs ARSD, EI, LCB, Hybrid and one-shot designs for  $n_0=9$ and $N=15$ over 50 simulations.}\label{figure_ex3.1}
\end{figure}

For Example \ref{ex3.1}, the initial sample size is set to $n_0 = 9$, with a total number of points $N = 15$. Figure \ref{figure_ex3.1} illustrates the average of the best minimum values obtained for different budget sizes for 50 simulations, demonstrating that the adaptive designs effectively identify the true minimum of the surface for $N=15$. Table \ref{Table_ex1_optim} summarizes the average of the best minimum values achieved using both one-shot and adaptive designs across 50 simulations. The results show that, with the addition of 6 points, adaptive designs successfully approach the true minimum value of -1.

\begin{table}[H]
\centering
\footnotesize
\caption{ \footnotesize The average of the best obtained minimum values of the response in Example \ref{ex3.1} for the adaptive designs ARSD, EI, LCB, Hybrid and one-shot designs for $n_0=9$ and $N=15$ over 50 simulations.}\label{Table_ex1_optim}
\resizebox{10cm}{!}{
\captionsetup[table]{skip=10pt}
\setlength\tabcolsep{5pt}
  \begin{tabular}{r rrrrrrrrr}
    \hline 
  $N$~~~\textbf{one-shot} & \textbf{ARSD}&\textbf{EI}& \textbf{LCB}&\textbf{Hybrid}&\\
   \hline
 \hline
10~~~~1.1092 & 0.2191 & -0.1034 & 0.2163 & -0.0014 \\ 
  11~~~~0.5139 & -0.3544 & -0.7418 & -0.3235 & -0.5389 \\ 
 12~~~~0.1751 & -0.7654 & -0.8898 & -0.7639 & -0.7766 \\ 
  13~~~-0.0596 & -0.9756 & -0.9422 & -0.9549 & -0.9555 \\ 
  14~~~-0.2236 & -0.9857 & -0.9630 & -0.9852 & -0.9871 \\ 
  15~~~-0.3805 & -0.9910 & -0.9794 & -0.9863 & -0.9962 \\ 
    \end{tabular}
    }
  \end{table}

\begin{figure}[H]
\centering
\includegraphics[scale=0.6]{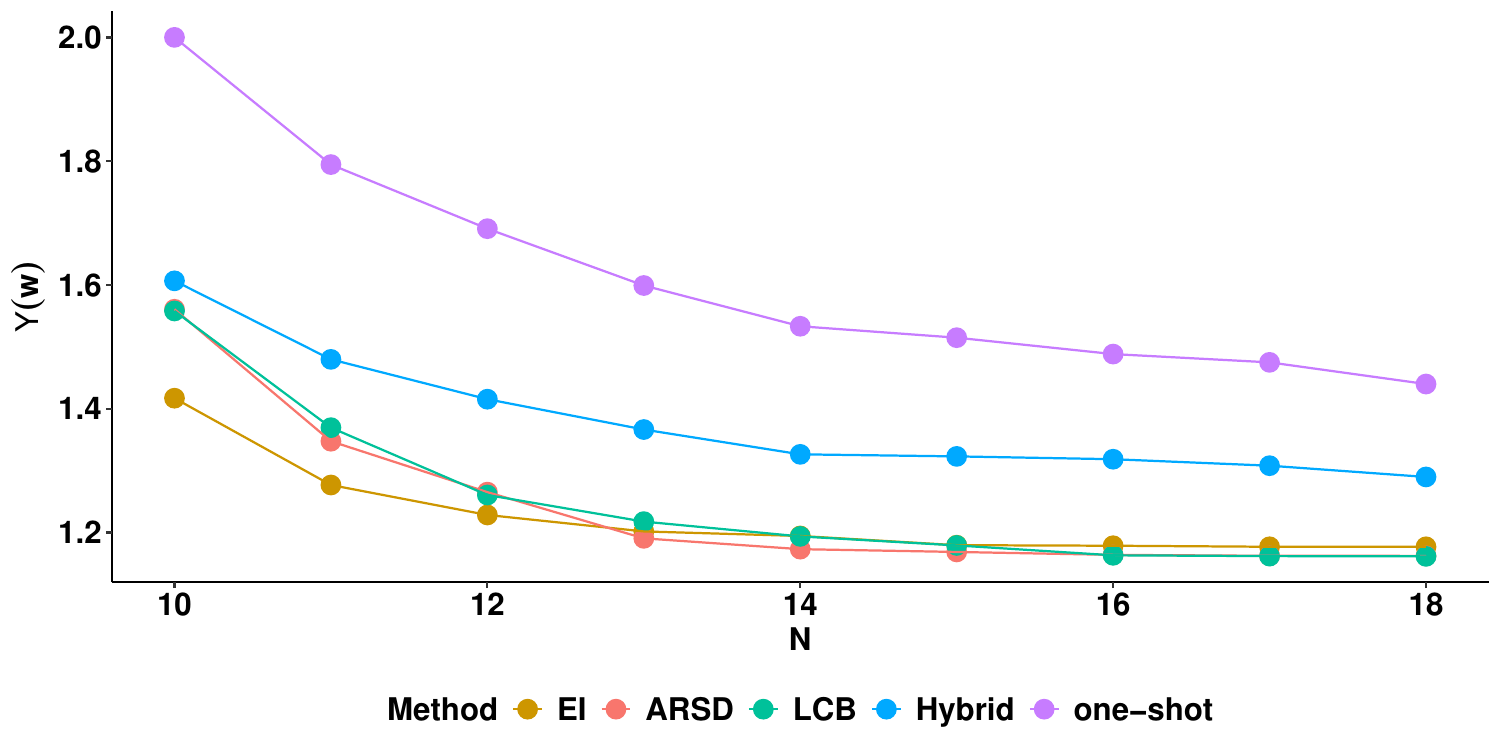}
\caption{\footnotesize The boxplot of obtained minimum values of the response in \nameref{ex3.2} for the adaptive designs ARSD, EI, LCB, Hybrid and one-shot designs for  $n_0=9$ and $N=18$ over 50 simulations.}\label{figure_ex3.2}
\end{figure}

For \nameref{ex3.2}, the initial sample size is set to $n_0 = 9$, with a total of $N = 18$ points. Figure \ref{figure_ex3.2} shows the average of the best minimum values obtained across 50 simulations for various budget sizes. The results demonstrate that adaptive designs consistently identify a more accurate true minimum of the surface. Among the adaptive methods, EI converges to the minimum faster than the others, while the Hybrid method takes the longest to converge. Table \ref{Table_ex2_optim} provides the average of the best minimum values achieved using both one-shot and adaptive designs across 50 simulations. The results indicate that ARSD and LCB methods yield minimum values closest to the true minimum for $N=18$.

\begin{table}[H]
\centering
\footnotesize
\caption{ \footnotesize The average of the best obtained minimum values of the response in \nameref{ex3.2} for the adaptive designs ARSD, EI, LCB, Hybrid and one-shot designs for $n_0=9$ and $N=18$ over 50 simulations.}\label{Table_ex2_optim}
\resizebox{10cm}{!}{
\captionsetup[table]{skip=10pt}
\setlength\tabcolsep{5pt}
  \begin{tabular}{r rrrrrrrrr}
    \hline 
  $N$~~~\textbf{one-shot} & \textbf{ARSD}&\textbf{EI}& \textbf{LCB}&\textbf{Hybrid}&\\
   \hline
 \hline

 10~~~~2.0002 & 1.5607 & 1.4170 & 1.5580 & 1.6067 \\ 
  11~~~~1.7946 & 1.3474 & 1.2767 & 1.3694 & 1.4797\\ 
  12~~~~1.6909 & 1.2652 & 1.2282 & 1.2605 & 1.4153 \\ 
  13~~~~1.5991 & 1.1905 & 1.2016 & 1.2175 & 1.3663 \\ 
  14~~~~1.5332 & 1.1730 & 1.1943 & 1.1936 & 1.3262 \\ 
  15~~~~1.5148 & 1.1685 & 1.1794 & 1.1790 & 1.3229 \\ 
  16~~~~1.4883 & 1.1637 & 1.1785 & 1.1630 & 1.3184 \\ 
  17~~~~1.4749 & 1.1622 & 1.1768 & 1.1617 & 1.3079 \\ 
  18~~~~1.4399 & 1.1621 & 1.1768 & 1.1613 & 1.2897 \\ 
    \end{tabular}
    }
  \end{table}
  
\begin{figure}[H]
\centering
\includegraphics[scale=0.6]{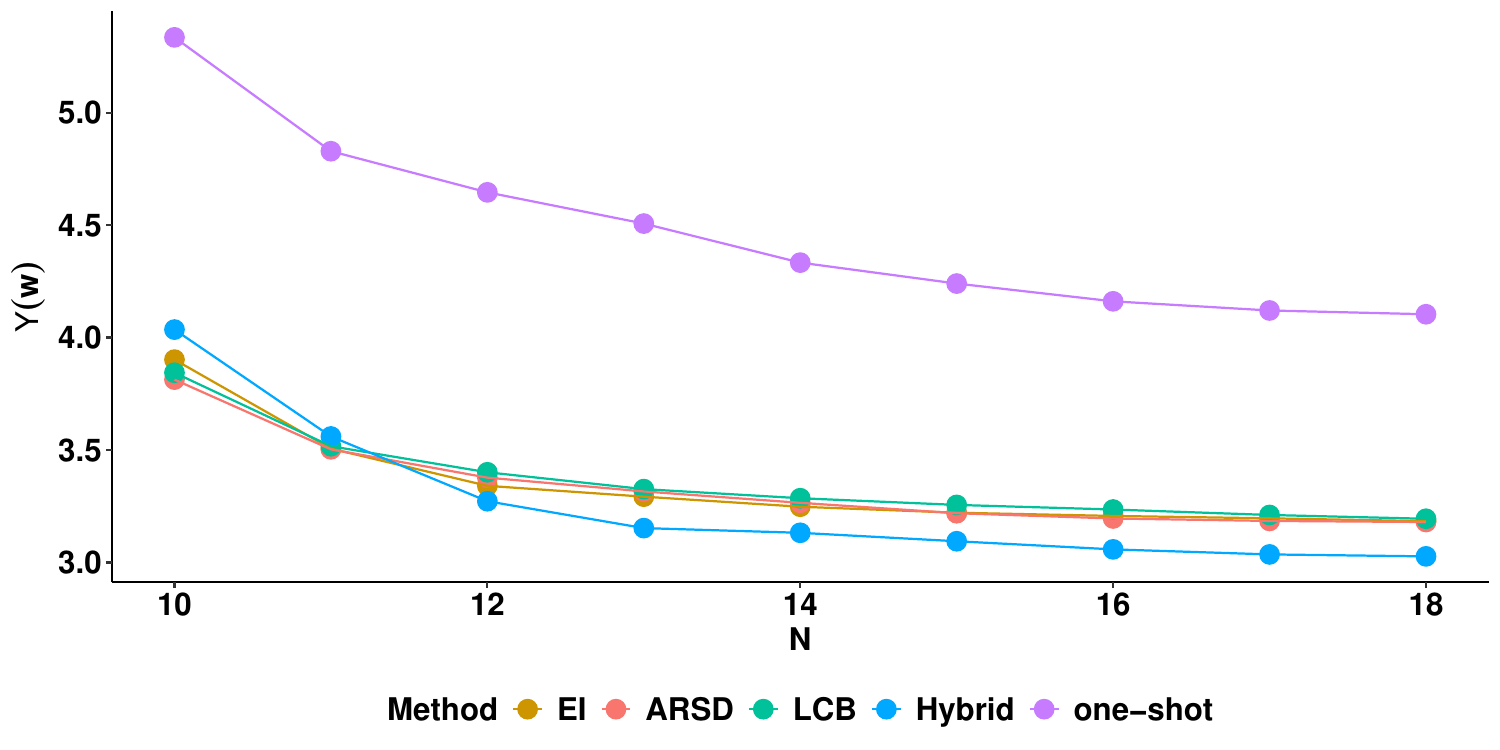}
\caption{\footnotesize The boxplot of obtained minimum values of the response in \nameref{ex3.3} for the adaptive designs ARSD, EI, LCB, Hybrid and one-shot designs for $n_0=9$ and $N=18$ over 50 simulations.}\label{figure_ex3.3}
\end{figure}

For \nameref{ex3.3}, the initial sample size is set to $n_0 = 9$, with a total of $N = 18$ points. Figure \ref{figure_ex3.3} illustrates the average of the best minimum values obtained across 50 simulations for various budget sizes. Similar to Example \ref{ex3.1} and \nameref{ex3.2}, the results show that adaptive designs consistently identify a more accurate true minimum of the surface. In this example, the Hybrid method converges to the minimum faster than the others and outperforms the other adaptive designs. Table \ref{Table_ex3_optim} provides the average of the best minimum values achieved using both one-shot and adaptive designs across 50 simulations. The results reveal that the Hybrid method successfully achieves a value close to the true minimum for $N = 18$.

\begin{table}[H]
\centering
\footnotesize
\caption{ \footnotesize The average of the best obtained minimum values of the response in \nameref{ex3.3} for the adaptive designs ARSD, EI, LCB, Hybrid and one-shot designs for $n_0=9$ and $N=18$ over 50 simulations.}\label{Table_ex3_optim}
\resizebox{10cm}{!}{
\captionsetup[table]{skip=10pt}
\setlength\tabcolsep{5pt}
  \begin{tabular}{r rrrrrrrrr}
    \hline 
  $N$~~~\textbf{one-shot} & \textbf{ARSD}&\textbf{EI}& \textbf{LCB}&\textbf{Hybrid}&\\
   \hline
 \hline
 10~~~~5.3346 & 3.8132 & 3.9019 & 3.8437 & 4.0355 \\ 
  11~~~~4.8285 & 3.5031 & 3.5061 & 3.5170 & 3.5603 \\ 
   12~~~~4.6456 & 3.3773 & 3.3408 & 3.4010 & 3.2724 \\ 
   13~~~~4.5068 & 3.3159 & 3.2926 & 3.3262 & 3.1529 \\ 
   14~~~~4.3331 & 3.2652 & 3.2479 & 3.2858 & 3.1320 \\ 
  15~~~~4.2401 & 3.2190 & 3.2212 & 3.2560 & 3.0945 \\ 
  16~~~~4.1610 & 3.1957 & 3.2070 & 3.2355 & 3.0583 \\ 
  17~~~~4.1203 & 3.1847 & 3.1964 & 3.2112 & 3.0359 \\ 
   18~~~~4.1036 & 3.1800 & 3.1837 & 3.1944 & 3.0273 \\ 

    \end{tabular}
    }
  \end{table}

\vspace{-1cm}
\subsection{Contour Estimation}\label{Section4.2}

In contour estimation, we call the EI for contour estimation (EI-C) to make it different from EI for optimization in (\ref{EIjones}). In our comparison, we consider RCC, EI-C, ECL, ARSD-C, LCB-C and one-shot designs discussed in Section \ref{Section3.2}.

To evaluate and compare the different approaches for contour estimation, we employ the following measurement $M_{C_{0}}$ used in \cite{Shahrokhian2024adaptive}, where
\begin{align}\label{mc0}
M_{C_{0}} = \frac{1}{|C_0|}\sum_{\bm{w}\in C_0} |Y(\bm{w}) - \hat{Y}(\bm{w})|,
\end{align}
where $C_0 = \{\bm{w}: Y(\bm{w}) = a\}$ is a subset of candidate points that satisfies $a \pm \epsilon$, with $\epsilon$ being a small positive value. As remarked by \cite{Shahrokhian2024adaptive}, the value of $\epsilon$ should be carefully chosen based on the complexity of the response function to ensure that a sufficient number of points are included.  To determine the size of the large set of candidate points used in $M_{C_{0}}$, we adopt the approach outlined by \cite{Shahrokhian2024adaptive}. In all examples, to calculate $M_{C_0}$, we generate 200-run LHDs for the quantitative inputs at each level combination of the qualitative inputs. Note that, we consider all possible combinations of the qualitative input levels.
It is important to note that the same dataset is utilized for approximating multiple contours in the EI-MC method described in (\ref{EIMC}).
 
For Example \ref{ex3.1}, the initial sample size is set to $n_0 = 9$, with a total number of points $N = \{11,13,15,17,19\}$. We examine contour levels $a = \{-0.7, 1.2,2.2\}$, with $\epsilon = 0.05$  for the $M_{C_0}$ measurement in (\ref{mc0}) and 
$\delta = 0.05$. Table \ref{Table_ex1} reports the average of
the $M_{C_0}$ values, obtained through both one-shot and adaptive designs across 50 simulations, demonstrate that adaptive designs provide more accurate contour estimations compared to one-shot designs. Although no single adaptive design method consistently outperforms the others, the RCC method often proves to be comparable to or superior to alternative approaches.
Table \ref{Ex1-time} presents the average computation time (in seconds) required to add 10 points adaptively. The results reveal that the ARSD-C and RCC methods are more efficient, requiring less computation time compared to the ECL, EI-C, and LCB-C methods.
\vspace{-0.2cm}
\begin{table}[H]
\centering
\footnotesize
\caption{\footnotesize  The average of the measurements $M_{C_{0}}$ in  Example \ref{ex3.1} for the adaptive designs ARSD-C, RCC, ECL, EI-C, LCB-C, and one-shot designs for $n_0=9$ and $N=\{11,13,15,17,19\}$ over 50 simulations for the contour level $a=\{-0.7,1.2,2.2\}$. The values in the parentheses are the relative efficiency of the criteria over one-shot designs.}\label{Table_ex1}
\resizebox{15cm}{!}{
\captionsetup[table]{skip=10pt}
\setlength\tabcolsep{2pt}
  \begin{tabular}{r rrrrrrrrr}
    \hline 
  $N$~~~\textbf{one-shot} & \textbf{ARSD-C} &\textbf{{RCC}}& \textbf{ECL} & \textbf{EI-C}& \textbf{LCB-C}&\\
   \hline
 \hline& &{a=-0.7}&&&\\
   \hline
11~~~~0.2361 & {0.0956(2.47) }& 0.1591(1.48)& 0.1959(1.21)& 0.2633(0.90)& \textbf{0.0893(2.64)}\\ 
 13~~~~0.1646 & 0.0360(4.57) &\textbf{0.0291(5.66)}& 0.0678(2.43)& 0.1517(1.09) & 0.0440(3.74)\\ 
 15~~~~0.0959 & \textbf{0.0022(43.5)}& 0.0052(18.4)& 0.0050 (19.1)& 0.0494(1.94)& {0.0075(12.8)}\\ 
17~~~~0.0823 & 0.0016(51.4)&{0.0002(412)}&\textbf{ 0.0001(823)}& 0.0039(21.1)& 0.0016(51.4) \\ 
 19~~~~0.0364 & \textbf{0.0001(364)} & \textbf{0.0001(364)}& \textbf{0.0001(364)}& 0.0004(91.0)& \textbf{0.0001(364)} \\ 
 
  \hline&&{a=1.2}&&&\\
   \hline
11~~~~0.4156 & 0.4598(0.90)& 0.4512(0.92)& 0.4478(0.93)& \textbf{0.4262(0.98)} & 0.4598(0.90)\\ 
13~~~~0.3434 & 0.3908(0.88)& \textbf{0.3625(0.95)}& 0.3869(0.89)& 0.3763(0.91)& 0.3899(0.88)\\ 
15~~~~0.2555 & 0.2514(1.02)& \textbf{0.2220(1.15)}& 0.2466(1.04)& 0.2350 (1.09)& 0.2536(1.01)\\ 
 17~~~~0.1580 & 0.1247(1.27)& 0.1300(1.22)& 0.1479(1.07)& \textbf{0.0887(1.78)}& 0.1251(1.26)\\ 
19~~~~0.0901 & 0.0298(3.02)& 0.0316(2.85)& 0.0295(3.05)& \textbf{0.0216(4.17)}& 0.0300(3.00)\\ 
   \hline& &{a=2.2}&&&\\
   \hline
11~~~~0.2846 & 0.1470(1.94) & \textbf{0.0239(11.9)} & 0.0404(7.04) & {0.1630(1.75)} & 0.1469(1.94) \\ 
13~~~~0.2275 & 0.0448(5.08) & 0.0061(37.3) & \textbf{0.0028(81.2)} & {0.1488(1.53)} & 0.0449(5.07) \\ 
15~~~~0.0540 & 0.0061(8.85) & \textbf{0.0008(67.5)} & 0.0009(60.0) & {0.0500(1.08)} & 0.0061(8.85) \\ 
17~~~~0.0266 & 0.0024(11.1) & 0.0005(53.2) & \textbf{0.0004(66.5)} & {0.0090(2.96)} & 0.0024(11.1) \\ 
19~~~~0.0096 & \textbf{0.0001(96.0)} & \textbf{0.0001(96.0)} & 0.0002(48.0) & {0.0008(12.0)} & \textbf{0.0001(96.0)} \\ 

    \end{tabular}
    }
  \end{table}

\begin{table}[H]
\footnotesize
   \centering
\caption{\footnotesize The average computation time (in seconds) in Example \ref{ex3.1} for the adaptive designs ARSD-C, RCC, ECL, EI-C, and LCB-C for $n_0=9$ and $N=19$ over 50 simulations for the contour level $a=\{-0.7,1.2,2.2\}$.}\label{Ex1-time}
\resizebox{9cm}{!}{
\captionsetup[table]{skip=10pt}
\setlength\tabcolsep{2pt}
  \begin{tabular}{r rrrrrrrrrrrr}
    \hline 
  a~~~~~&&\textbf{ARSD-C}& &\textbf{{RCC}}&  &\textbf{ECL} & &\textbf{EI-C}& &\textbf{LCB-C}\\[0.05cm]  \hline
   -0.7~~~~~&&0.14& &0.13&  &0.77& &0.78& &0.79\\
   1.2~~~~~&&0.22& &0.49& &0.80 & &0.77& &0.75\\
   2.2~~~~~&&0.17& &0.15& &0.73 & &0.72& &0.73\\
      \end{tabular}
      }
  \end{table}

For \nameref{ex3.2}, the contour levels of interest are set to  $a = \{1.2, 1.7, 2.1\}$. The initial sample size is  $n_0 = 9$, with total points $N = \{27, 36, 45, 54,63\}$. We use $\epsilon = 0.05$ for the 
$M_{C_0}$ 	
 measurement in (\ref{mc0}) and $\delta=0.02$. Table \ref{Table_ex2} provides a summary of the average $M_{C_0}$ 	
  values obtained using one-shot and adaptive designs across 50 simulations for the specified contour levels and run sizes. The results indicate that the RCC method consistently exhibits superior relative efficiency for different contour levels and most of the run sizes, outperforming the other criteria in the comparison.
In addition, Table \ref{Ex2-time} presents the average computation time (in seconds) required to add 54 points adaptively. The findings demonstrate that the ARSD-C and RCC methods are significantly faster than the ECL, EI-C, and LCB-C methods, further emphasizing their computational efficiency.

\begin{table}[H]
    \centering
\caption{\footnotesize  The average of the measurements $M_{C_{0}}$ in \nameref{ex3.2} for the adaptive designs ARSD-C, RCC, ECL, EI-C, LCB-C and one-shot designs for $n_0=9$ and $N=\{27,36,45,54,63\}$ over 50 simulations for the contour level $a=\{1.2,1.7,2.1\}$. The values in the parentheses are the relative efficiency of the criteria over one-shot designs.}\label{Table_ex2}
\resizebox{15cm}{!}{
\captionsetup[table]{skip=10pt}
\setlength\tabcolsep{2pt}
  \begin{tabular}{r rrrrrrrrr}
 
  \hline
  $N$~~~\textbf{one-shot} & \textbf{ARSD-C} &\textbf{{RCC}}& \textbf{ECL} & \textbf{EI-C}& \textbf{LCB-C}&\\[0.05cm]
   \hline
 \hline& &{a=1.2}&&&\\
   \hline
 27~~~~0.0824 & 0.0139(5.93) & 0.0081(10.17) & \textbf{0.0070(11.7)} & 0.0247(3.34) & 0.0134(6.15) \\ 
 36~~~~0.0335 & 0.0016(20.9) & 0.0010(33.5) & \textbf{0.0008(41.8)} & 0.0050(6.70) & 0.0023(14.5) \\ 
 45~~~~0.0157 & 0.0003(52.3) & \textbf{0.0002(78.5)} & 0.0003(52.3) & 0.0005(31.4) & 0.0003(52.3) \\ 
 54~~~~0.0073 & \textbf{0.0001(73.0)} & \textbf{0.0001(73.0)} & \textbf{0.0001(73.0)} & 0.0002(36.5)&\textbf{0.0001(73.0)} \\ 
 63~~~~0.0045 & \textbf{0.0001(45.0)} & \textbf{0.0001(45.0)} & \textbf{0.0001(45.0)} & 0.0002(22.5) & \textbf{0.0001(45.0)} \\ 
    \hline& &{a=1.7}&&&\\
   \hline
27~~~~0.0643 & \textbf{0.0500(1.29)} & 0.0555(1.16) & {0.0577(1.11)} & 0.0556(1.16) & {0.0514(1.25)} \\ 
36~~~~0.0214 & 0.0170(1.26) & \textbf{0.0147(1.46)} & 0.0167(1.28) & 0.0175(1.22) & 0.0156(1.37) \\ 
45~~~~0.0115 & 0.0034(3.38) & \textbf{0.0031(3.71)} & \textbf{0.0031(3.71)} & 0.0045(2.56) & 0.0037(3.11) \\ 
54~~~~0.0042 & 0.0009(4.67) & \textbf{0.0009(4.67)} & {0.0010(4.20)} & 0.0011(3.82) & 0.0010(4.20) \\ 
63~~~~0.0020 & \textbf{0.0004(5.00)} & \textbf{0.0004(5.00)} & {0.0005(4.00)} & \textbf{0.0004(5.00)} & \textbf{0.0004(5.00)} \\ 
 \hline& &{a=2.1}&&&\\
 \hline
27~~~~0.0663 & 0.0528(1.26) & 0.0553(1.20) & \textbf{0.0503(1.32)} & 0.0563(1.18) & 0.0540(1.23) \\ 
36~~~~0.0239 & 0.0160(1.49) & 0.0197(1.21) & \textbf{0.0141(1.70)} & 0.0200(1.20) & 0.0175(1.37) \\ 
45~~~~0.0094 & 0.0038(2.47) & \textbf{0.0034(2.76)} & \textbf{0.0034(2.76)} & 0.0046(2.04) & 0.0039(2.41) \\ 
54~~~~0.0040 & 0.0012(3.33) & \textbf{0.0011(3.64)} & {0.0013(3.08)} & 0.0012(3.33) & 0.0011(3.64) \\ 
63~~~~0.0018 & 0.0005(3.60) & \textbf{0.0004(4.50)} & {0.0005(3.60)} & 0.0006(3.00) & 0.0005(3.60) \\ 

    \end{tabular}
   }
  \end{table}

 \begin{table}[H]
\footnotesize
   \centering
\caption{\footnotesize The average computation time per second in \nameref{ex3.2} for the adaptive designs ARSD-C, RCC,  ECL, EI-C, and LCB-C  for $n_0=9$ and $N=63$ over 50 simulations for the contour level $a=\{1.2,1.7,2.1\}$.}  \label{Ex2-time}
\resizebox{9cm}{!}{
\captionsetup[table]{skip=10pt}
\setlength\tabcolsep{2pt}
  \begin{tabular}{r rrrrrrrrrrrr}
    \hline 
  $a$~~~~~&&\textbf{ARSD-C}& &\textbf{{RCC}}& &\textbf{ECL} & &\textbf{EI-C}& &\textbf{LCB-C}\\[0.05cm]  \hline
   1.2~~~~~&&7.14& &7.77& &34.8 & &34.3& &34.9\\
   1.7~~~~~&&6.49& &7.28& &34.3 & &36.2& &35.4\\
   2.1~~~~~&&8.14& &8.18& &34.8 & &36.3& &35.3\\
\end{tabular}
      }
  \end{table}
 
For \nameref{ex3.3}, the contour levels of interest are set as $a = \{4.5, 5.5, 6.5\}$. We use an initial sample size of $n_0 = 9$, with total points $N = \{27, 36, 45, 54,63\}$. The measurement $M_{C_0}$ in (\ref{mc0}) is evaluated using $\epsilon = 0.1$, and we set $\delta = 0.1$. Table \ref{Table_ex3} presents the average $M_{C_0}$ values obtained from one-shot designs and adaptive designs over 50 simulations. Table \ref{Ex3-time} provides the average computation time required to add 54 points adaptively for each adaptive design method. The results are consistent with those observed in Example \ref{ex3.1} and \nameref{ex3.2}, demonstrating similar trends in performance and efficiency.

\begin{table}[H]
    \centering
\caption{\footnotesize  The average of the measurements $M_{C_{0}}$ in \nameref{ex3.3} for the adaptive designs ARSD-C, RCC, ECL, EI-C, LCB-C and one-shot designs for $n_0=9$ and $N=\{27,36,45,54,63\}$ over 50 simulations for the contour level $a=\{4.5,5.5,6.5\}$. The values in the parentheses are the relative efficiency of the criteria over one-shot designs.}\label{Table_ex3}
\resizebox{15cm}{!}{
\captionsetup[table]{skip=10pt}
\setlength\tabcolsep{2pt}
  \begin{tabular}{r rrrrrrrrr}
    \hline 
  $N$~~~\textbf{one-shot} & \textbf{ARSD-C} &\textbf{{RCC}}& \textbf{ECL} & \textbf{EI-C}& \textbf{LCB-C}&\\[0.05cm]
   \hline
 \hline& &{a=4.5}&&&\\
   \hline

 27~~~~0.1671 & \textbf{0.1232(1.36)} & 0.1412(1.18) & {0.1434(1.17)} & 0.1353(1.24) & 0.1248(1.34) \\ 
 36~~~~0.1256 & 0.1040(1.21) & 0.1028(1.22) & \textbf{0.0980(1.28)} & 0.1088(1.15) & 0.1028(1.22) \\ 
 45~~~~0.0975 & 0.0889(1.10) & \textbf{0.0714(1.37)} & 0.0719(1.36) & 0.0947(1.03) & 0.0886(1.10) \\ 
 54~~~~0.0709 & 0.0619(1.15) & \textbf{0.0374(1.90)} & 0.0414(1.71) & 0.0724(0.98) & 0.0642(1.10) \\ 
 63~~~~0.0384 & 0.0293(1.31) & {0.0227(1.69)} & \textbf{0.0207(1.86)} & 0.0383(1.00) & 0.0300(1.28) \\ 
    \hline& &{a=5.5}&&\\
   \hline
 27~~~~0.1453 & \textbf{0.1215(1.20)} & 0.1265(1.15) & {0.1296(1.12)} & 0.1224(1.19) & 0.1177(1.23) \\ 
 36~~~~0.1144 & 0.1016(1.13) & \textbf{0.0967(1.18)} & 0.0979(1.17) & 0.1036(1.10) & 0.1029(1.11) \\ 
 45~~~~0.0908 & 0.0874(1.04) & {0.0745(1.22)} & \textbf{0.0713(1.27)} & 0.0845(1.07) & 0.0814(1.12) \\ 
 54~~~~0.0643 & 0.0531(1.21) & \textbf{0.0406(1.58)} & 0.0421(1.53) & 0.0600(1.07) & 0.0494(1.30) \\ 
63~~~~0.0338 & 0.0216(1.56) & {0.0202(1.67)} & \textbf{0.0200(1.69)} & 0.0251(1.35) & 0.0201(1.68) \\ 
 \hline& &{a=6.5}&&\\
 \hline
27~~~~0.3184 & 0.2901(1.10) & 0.3487(0.91) & {0.3521(0.90)} & \textbf{0.2728(1.17)}& 0.2978(1.07) \\ 
36~~~~0.2623 & \textbf{0.1175(2.23)} & 0.1606(1.63) & {0.1787(1.47)} & 0.1403(1.87) & 0.1624(1.62) \\ 
45~~~~0.2142 & 0.0592(3.62) & \textbf{0.0414(5.17)} & {0.0523(4.10)} & 0.0444(4.82) & 0.0644(3.33) \\ 
54~~~~0.1463 & 0.0190(7.70) & \textbf{0.0088(16.6)} & {0.0134(10.9)} & 0.0138(10.6) & 0.0158(9.26) \\ 
63~~~~0.0863 & 0.0117(7.38) & 0.0038(22.7) & {0.0067(12.8)} & 0.0040(21.6) & \textbf{0.0030(28.7)} \\ 
\end{tabular}
}
  \end{table}

  \begin{table}[H]
\footnotesize
   \centering
\caption{\footnotesize The average computation time per second in \nameref{ex3.3} for the adaptive designs ARSD-C, RCC, ECL, EI-C, and LCB-C for $n_0=9$ and $N=63$ over 50 simulations for the contour level $a=\{4.5,5.5,6.5\}$.}\label{Ex3-time}
\resizebox{9cm}{!}{
\captionsetup[table]{skip=10pt}
\setlength\tabcolsep{2pt}
  \begin{tabular}{r rrrrrrrrrrrr}
    \hline 
  $a$~~~~~&&\textbf{ARSD-C}& &\textbf{{RCC}}& &\textbf{ECL} & &\textbf{EI-C}& &\textbf{LCB-C}\\[0.05cm]  \hline
   4.5~~~~~&&57.7& &55.6& &116.8& &116.8& &118.5\\
   5.5~~~~~&&69.1& &63.5&  &116.9 & &107.3& &117.7\\
   6.5~~~~~&&26.7& &24.3& &123.6 & &120.7& &117.9\\
      \end{tabular}
      }
  \end{table}

\subsection{Prediction}\label{Section4.3}

This subsection applies the criteria EI-MC and EI-SC (discussed in Section \ref{Section2.2}) that were originally proposed for the problem of prediction for computer experiments with only quantitative inputs to the problem of prediction for computer experiments 
with both quantitative and qualitative inputs. 

For the initial run size and budget size, we follow the recommendations provided by \cite{yang2020global}.
The initial sample size is set to be \( n_0 = 5(p+q) \) the maximum budget is set to be \( 20(p+q) \). For EI-MC we set $c$ for 10 multiple contours. Similar to optimization and  contour estimation the next input is selected by evaluating a large number of candidate points in the design space.

\begin{figure}[H]
\centering
\includegraphics[scale=0.4]{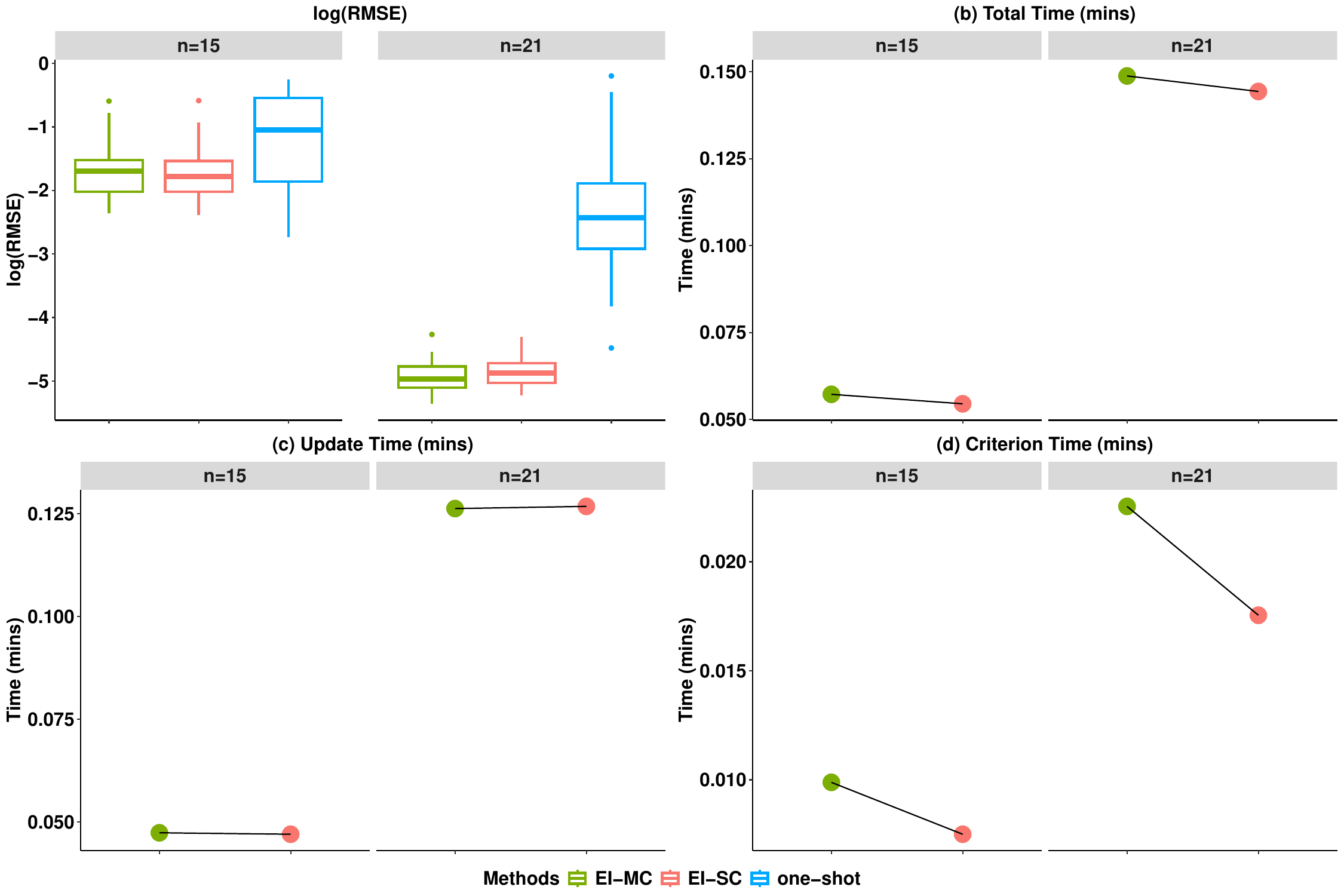}
\caption{\footnotesize (a) The boxplot of the log of RMSEs; (b) The average of total time in minutes; (c) The average of update time in minutes;  (d) The average of criterion time in minutes;  for  $n_0=10$ and $N=30$ over 50 simulations in Example \ref{ex3.1}.}\label{figure1_ex3.1}
\end{figure}

\begin{figure}[H]
\centering
\includegraphics[scale=0.4]{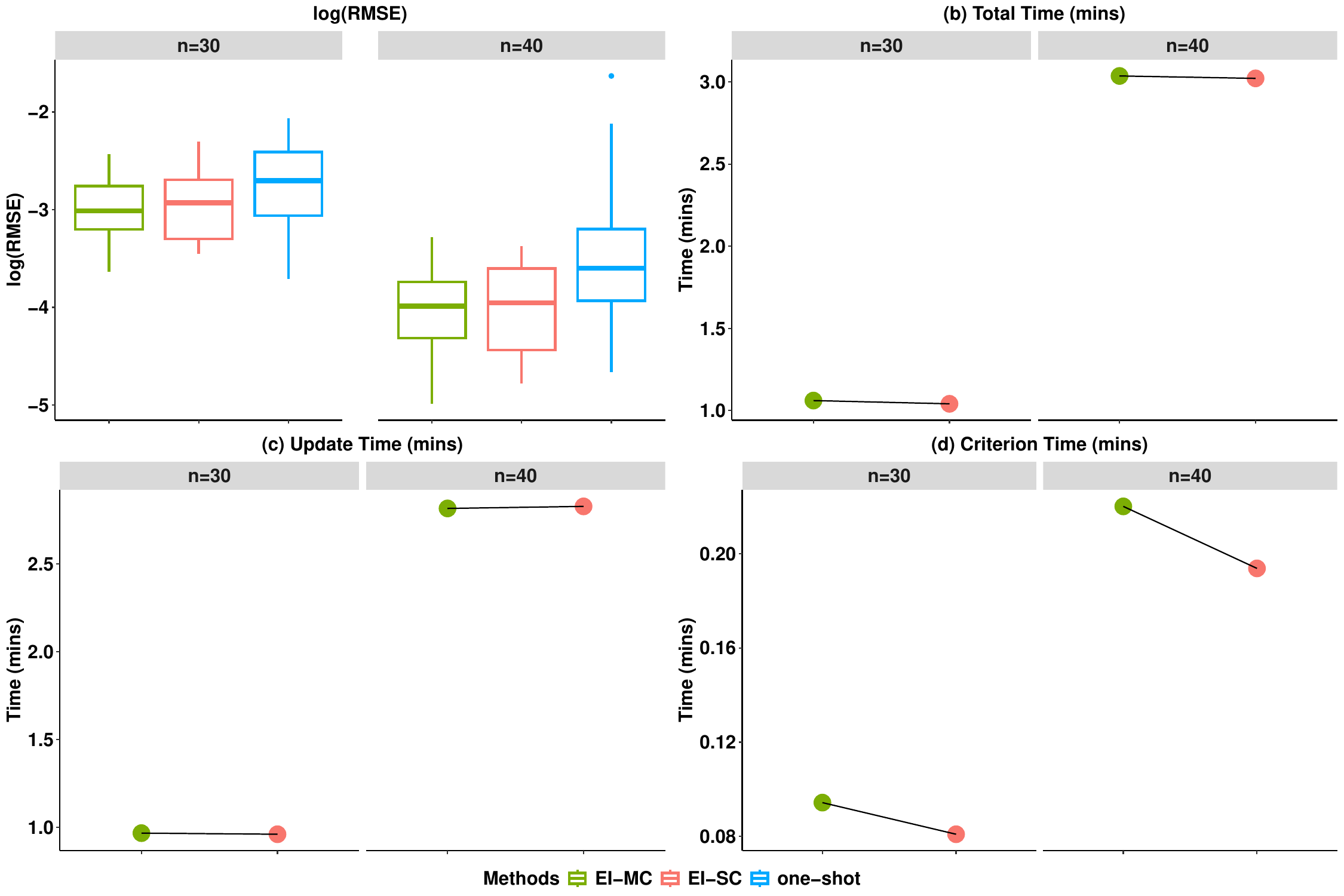}
\caption{\footnotesize (a) The boxplot of the log of RMSEs; (b) The average of total time in minutes; (c) The average of update time in minutes;  (d) The average of criterion time in minutes;  for  $n_0=20$ and $N=40$ over 50 simulations in \nameref{ex3.2}.}\label{figure1_ex3.2}
\end{figure}

\begin{figure}[H]
\centering
\includegraphics[scale=0.4]{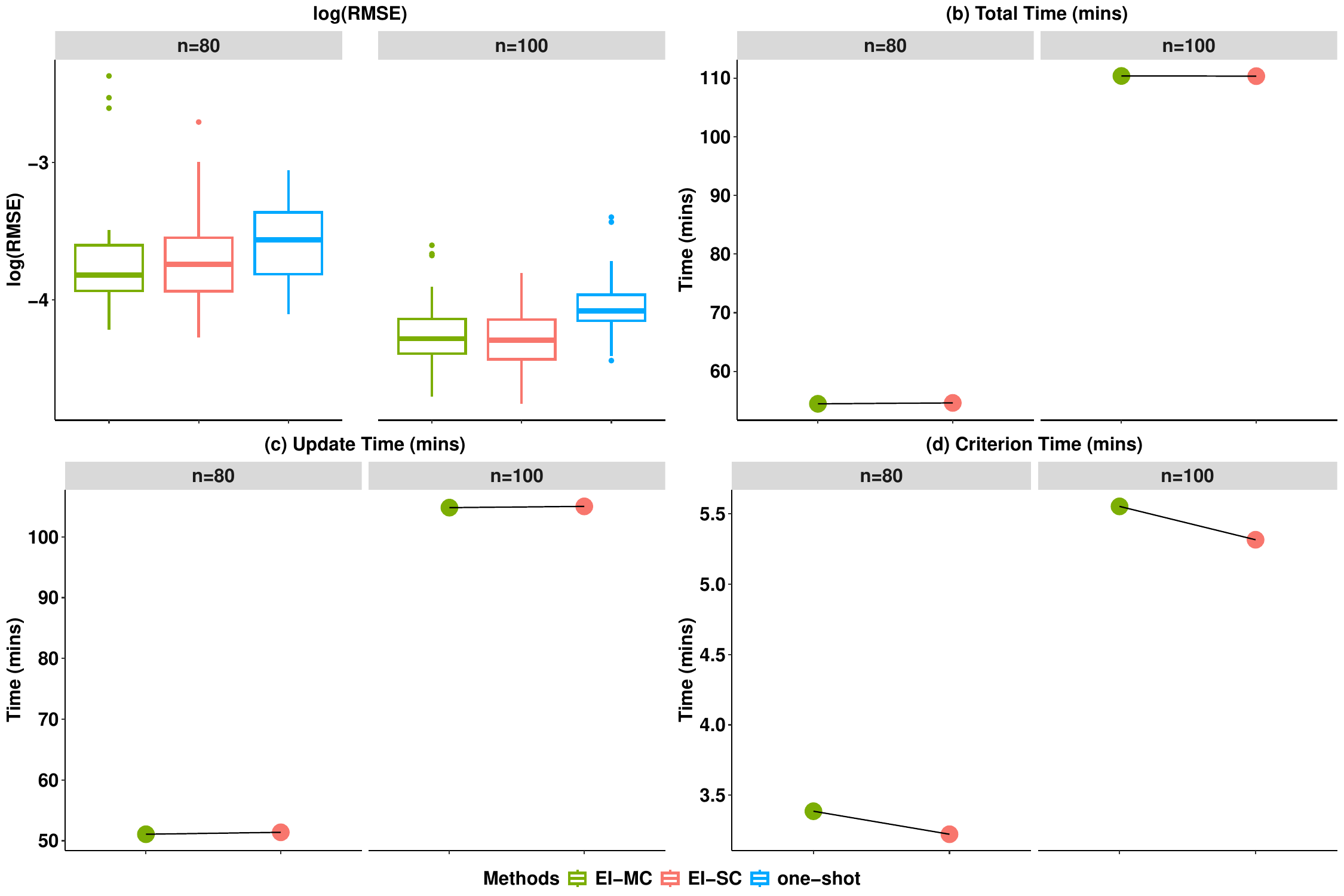}
\caption{\footnotesize (a) The boxplot of the log of RMSEs; (b) The average of total time in minutes; (c) The average of update time in minutes;  (d) The average of criterion time in minutes;  for  $n_0=30$ and $N=100$ over 50 simulations in \nameref{ex3.3}.}\label{figure1_ex3.3}
\end{figure}

To evaluate prediction accuracy, the predicted function is compared with the true function using the root mean square error (RMSE), defined as
\begin{align}\label{RMSE}
\mbox{RMSE} = \sqrt{\frac{1}{n_w}\sum_{i=1}^{n_w} (\hat{Y}(\bm{w}^{*}_i) - Y(\bm{w}^{*}_i))^2},
\end{align}
where $n_w$ is the number of data points in the test set. In the captions of the figures, the values in parentheses represent the total size of the data for each method, and in the figures, we present the logarithm of the RMSE (base \( e \)). We record the computation time in minutes for adaptive designs EI-MC and EI-SC. Figure \ref{figure1_ex3.1} to Figure \ref{figure1_ex3.3} each present four panels. Panel (a) displays the logarithm of the RMSE in (\ref{RMSE}), to evaluate prediction accuracy for each adaptive design. Panel (b) shows the average total time required for the adaptive design to sequentially add points and update the model. Panel (c) provides the time each taken to update the model. Panel (d) illustrates the time needed to select the next input in each adaptive design.

For Example \ref{ex3.1} with 
$n_0 = 10$ and $N=\{15,21\}$, Figure~\ref{figure1_ex3.1} demonstrates that adaptive designs achieve significantly better prediction accuracy than one-shot designs. However, the total computational time for EI-MC is higher than EI-SC, as observed in panel (b). This discrepancy stems from the time-intensive calculation of the EI-MC criterion, which requires evaluating EI across 10 contours. 

In  \nameref{ex3.2} with $n_0 = 20$ and  $N=\{30,40\}$, Figure~\ref{figure1_ex3.2} similarly shows that adaptive designs outperform one-shot designs. However, the computation time increases substantially due to the higher dimensionality, larger initial design size, and expanded budget. The analysis is capped at $N=40$, as the advantage of adaptive designs becomes apparent by this point.
 
For  \nameref{ex3.3} with $n_0 = 30$ and  $N=\{80,100\}$, Figure~\ref{figure1_ex3.3} confirms that adaptive designs consistently deliver more accurate predictions than one-shot designs. The computation time for model updates in this case is notably higher than in the previous examples, reflecting the increased complexity of modeling computer experiments with mixed inputs, which involves estimating a larger number of parameters.

\section{Concluding Remarks}\label{Section5}

Adaptive designs are an iterative and goal-oriented approach to selecting inputs for computer experiments.  They dynamically update the emulators based on the data at hand, improving efficiency and accuracy. These designs are particularly beneficial in scenarios where computational resources are limited or objectives such as optimization, contour estimation, or prediction need to be achieved with minimal computational cost.  They continue to evolve with innovations in modeling and computation, particularly for computer experiments with both quantitative and qualitative input variables. 
In this work, we review and discuss various adaptive designs for optimization, contour estimation, and prediction in computer experiments involving mixed inputs.

We highlight several key issues that are worth further investigation in computer experiments involving mixed inputs. The efficiency of an adaptive design relies heavily on a well-structured surrogate model and an effective criterion to achieve the desired objectives. While this study utilized the EzGP model as the surrogate, alternative models could also be explored. However, a significant challenge with models for mixed inputs is the computational cost, as they often require estimating a larger number of parameters. This challenge becomes particularly pronounced in high-dimensional settings, calling for cost-effective modeling approaches tailored to mixed inputs.

Another direction for future research would be adaptive designs for prediction when the model involves both quantitative and qualitative inputs.  We applied the criteria originally proposed for prediction in computer experiments with only quantitative inputs to computer experiments with mixed inputs. Although the results showed that such adaptive designs outperform one-shot designs.  To the best of our knowledge, no specific work has been done on adaptive designs for prediction in computer experiments with both quantitative and qualitative inputs, and this area remains open for exploration.

In addition, for models with mixed inputs, it is not straightforward to directly identify the inputs that optimize an adaptive design criterion. In this work, we addressed this limitation by evaluating a large set of candidate points to approximate the optimal solution. However, this approach is computationally intensive and represents an open and critical challenge for further research. Developing more efficient methods to identify optimal inputs in adaptive designs for mixed inputs remains an important area for future exploration.

\bibliography{references}

\begin{thebibliography}{37}
\providecommand{\natexlab}[1]{#1}
\providecommand{\url}[1]{\texttt{#1}}
\expandafter\ifx\csname urlstyle\endcsname\relax
  \providecommand{\doi}[1]{doi: #1}\else
  \providecommand{\doi}{doi: \begingroup \urlstyle{rm}\Url}\fi

\bibitem[Auer(2002)]{auer2002using}
Peter Auer.
\newblock Using confidence bounds for exploitation-exploration trade-offs.
\newblock \emph{Journal of Machine Learning Research}, 3:\penalty0 397--422, 2002.

\bibitem[Bect et~al.(2012)Bect, Ginsbourger, Li, Picheny, and Vazquez]{bect2012sequential}
Julien Bect, David Ginsbourger, Ling Li, Victor Picheny, and Emmanuel Vazquez.
\newblock Sequential design of computer experiments for the estimation of a probability of failure.
\newblock \emph{Statistics and Computing}, 22:\penalty0 773--793, 2012.

\bibitem[Cai et~al.(2024)Cai, Xu, Lin, Hong, and Deng]{cai2024adaptive}
Xia Cai, Li~Xu, C~Devon Lin, Yili Hong, and Xinwei Deng.
\newblock Adaptive-region sequential design with quantitative and qualitative factors in application to {HPC} configuration.
\newblock \emph{Journal of Quality Technology}, 56\penalty0 (1):\penalty0 5--19, 2024.

\bibitem[Cameron et~al.(2019)Cameron, Anwar, Cheng, Xu, Li, Ananth, Bernard, Jearls, Lux, Hong, et~al.]{cameron2019moana}
Kirk~W Cameron, Ali Anwar, Yue Cheng, Li~Xu, Bo~Li, Uday Ananth, Jon Bernard, Chandler Jearls, Thomas Lux, Yili Hong, et~al.
\newblock Moana: Modeling and analyzing {I/O} variability in parallel system experimental design.
\newblock \emph{IEEE Transactions on Parallel and Distributed Systems}, 30\penalty0 (8):\penalty0 1843--1856, 2019.

\bibitem[Carnell(2022)]{r-core}
Rob Carnell.
\newblock \emph{Latin Hypercube Samples}.
\newblock R Foundation for Statistical Computing, 2022.
\newblock URL \url{https://cran.r-project.org/package=lhs}.

\bibitem[Cole et~al.(2023)Cole, Gramacy, Warner, Bomarito, Leser, and Leser]{cole2023entropy}
D~Austin Cole, Robert~B Gramacy, James~E Warner, Geoffrey~F Bomarito, Patrick~E Leser, and William~P Leser.
\newblock Entropy-based adaptive design for contour finding and estimating reliability.
\newblock \emph{Journal of Quality Technology}, 55\penalty0 (1):\penalty0 43--60, 2023.

\bibitem[Deng et~al.(2017)Deng, Lin, Liu, and Rowe]{deng2017additive}
X~Deng, C~Devon Lin, K-W Liu, and RK~Rowe.
\newblock Additive {Gaussian} process for computer models with qualitative and quantitative factors.
\newblock \emph{Technometrics}, 59\penalty0 (3):\penalty0 283--292, 2017.

\bibitem[Deng et~al.(2015)Deng, Hung, and Lin]{deng2015design}
Xinwei Deng, Ying Hung, and C~Devon Lin.
\newblock Design for computer experiments with qualitative and quantitative factors.
\newblock \emph{Statistica Sinica}, 25:\penalty0 1567--1581, 2015.

\bibitem[Gramacy(2020)]{gramacy2020surrogates}
Robert~B Gramacy.
\newblock \emph{Surrogates: {Gaussian} process modeling, design, and optimization for the applied sciences}.
\newblock Chapman and Hall/CRC, 2020.

\bibitem[Han et~al.(2009)Han, Santner, Notz, and Bartel]{han2009prediction}
Gang Han, Thomas~J Santner, William~I Notz, and Donald~L Bartel.
\newblock Prediction for computer experiments having quantitative and qualitative input variables.
\newblock \emph{Technometrics}, 51\penalty0 (3):\penalty0 278--288, 2009.

\bibitem[He et~al.(2017{\natexlab{a}})He, Lin, and Sun]{he2017construction}
Yuanzhen He, C~Devon Lin, and Fasheng Sun.
\newblock On construction of marginality coupled designs.
\newblock \emph{Statistica Sinica}, 27:\penalty0 665--683, 2017{\natexlab{a}}.

\bibitem[He et~al.(2017{\natexlab{b}})He, Lin, Sun, and Lv]{he2017marginally}
Yuanzhen He, C~Devon Lin, Fasheng Sun, and Benjian Lv.
\newblock Marginally coupled designs for two-level qualitative factors.
\newblock \emph{Journal of Statistical Planning and Inference}, 187:\penalty0 103--108, 2017{\natexlab{b}}.

\bibitem[Hern{\'a}ndez-Lobato et~al.(2014)Hern{\'a}ndez-Lobato, Hoffman, and Ghahramani]{hernandez2014predictive}
Jos{\'e}~Miguel Hern{\'a}ndez-Lobato, Matthew~W Hoffman, and Zoubin Ghahramani.
\newblock Predictive entropy search for efficient global optimization of black-box functions.
\newblock \emph{Advances in Neural Information Processing Systems}, 27, 2014.

\bibitem[Hwang et~al.(2016)Hwang, He, and Qian]{hwang2016sliced}
Youngdeok Hwang, Xu~He, and Peter~ZG Qian.
\newblock Sliced orthogonal array-based {Latin} hypercube designs.
\newblock \emph{Technometrics}, 58\penalty0 (1):\penalty0 50--61, 2016.

\bibitem[Jones et~al.(1998)Jones, Schonlau, and Welch]{jones1998efficient}
Donald~R Jones, Matthias Schonlau, and William~J Welch.
\newblock Efficient global optimization of expensive black-box functions.
\newblock \emph{Journal of Global Optimization}, 13\penalty0 (4):\penalty0 455--492, 1998.

\bibitem[Kocsis and Szepesv{\'a}ri(2006)]{kocsis2006bandit}
Levente Kocsis and Csaba Szepesv{\'a}ri.
\newblock Bandit based {Monte}-{Carlo} planning.
\newblock \emph{European Conference on Machine Learning}, 282--293, 2006.

\bibitem[Li et~al.(2023)Li, Xiao, Mandal, Lin, and Deng]{r-EzGP}
Jiayi Li, Qian Xiao, Abhyuday Mandal, C.~Devon Lin, and Xinwei Deng.
\newblock \emph{Easy-to-Interpret {Gaussian} Process Models for Computer Experiments with Both Quantitative and Qualitative Factors}.
\newblock R Foundation for Statistical Computing, 2023.
\newblock URL \url{https://cran.r-project.org/package=EzGP}.

\bibitem[Lin et~al.(2024)Lin, Sung, and Chen]{lin2024category}
Wei-Ann Lin, Chih-Li Sung, and Ray-Bing Chen.
\newblock Category tree {Gaussian} process for computer experiments with many-category qualitative factors and application to cooling system design.
\newblock \emph{Journal of Quality Technology}, 56\penalty0 (5):\penalty0 391--408, 2024.

\bibitem[Luo et~al.(2024)Luo, Cho, Demmel, Li, and Liu]{luo2024hybrid}
Hengrui Luo, Younghyun Cho, James~W Demmel, Xiaoye~S Li, and Yang Liu.
\newblock Hybrid parameter search and dynamic model selection for mixed-variable {Bayesian} optimization.
\newblock \emph{Journal of Computational and Graphical Statistics}, 1--14, 2024.

\bibitem[McKay et~al.(1979)McKay, Beckman, and Conover]{mckay2000comparison}
Michael~D McKay, Richard~J Beckman, and William~J Conover.
\newblock A comparison of three methods for selecting values of input variables in the analysis of output from a computer code.
\newblock \emph{Technometrics}, 42\penalty0 (1):\penalty0 55--61, 1979.

\bibitem[Mebane et~al.(2011)Mebane, Sekhon, and Saarinen]{r-genoud}
Walter~R Mebane, Jasjeet~Singh Sekhon, and Theo Saarinen.
\newblock \emph{Genetic Optimization Using Derivatives}.
\newblock R Foundation for Statistical Computing, 2011.
\newblock URL \url{https://cran.r-project.org/package=rgenoud}.

\bibitem[Mohammadi et~al.(2022)Mohammadi, Challenor, Williamson, and Goodfellow]{mohammadi2022cross}
Hossein Mohammadi, Peter Challenor, Daniel Williamson, and Marc Goodfellow.
\newblock Cross-validation--based adaptive sampling for {Gaussian} process models.
\newblock \emph{SIAM/ASA Journal on Uncertainty Quantification}, 10\penalty0 (1):\penalty0 294--316, 2022.

\bibitem[Picheny et~al.(2013)Picheny, Ginsbourger, Richet, and Caplin]{picheny2013quantile}
Victor Picheny, David Ginsbourger, Yann Richet, and Gregory Caplin.
\newblock Quantile-based optimization of noisy computer experiments with tunable precision.
\newblock \emph{Technometrics}, 55\penalty0 (1):\penalty0 2--13, 2013.

\bibitem[Qian et~al.(2008)Qian, Wu, and Wu]{qian2008gaussian}
Peter Z~G Qian, Huaiqing Wu, and CF~Jeff Wu.
\newblock {Gaussian} process models for computer experiments with qualitative and quantitative factors.
\newblock \emph{Technometrics}, 50\penalty0 (3):\penalty0 383--396, 2008.

\bibitem[Qian(2012)]{qian2012sliced}
Peter~ZG Qian.
\newblock Sliced {Latin} hypercube designs.
\newblock \emph{Journal of the American Statistical Association}, 107\penalty0 (497):\penalty0 393--399, 2012.

\bibitem[Ranjan et~al.(2008)Ranjan, Bingham, and Michailidis]{ranjan2008sequential}
Pritam Ranjan, Derek Bingham, and George Michailidis.
\newblock Sequential experiment design for contour estimation from complex computer codes.
\newblock \emph{Technometrics}, 50\penalty0 (4):\penalty0 527--541, 2008.

\bibitem[Sacks et~al.(1989)Sacks, Welch, Mitchell, and Wynn]{sacks1989design}
Jerome Sacks, William~J Welch, Toby~J Mitchell, and Henry~P Wynn.
\newblock Design and analysis of computer experiments.
\newblock \emph{Statistical Science}, 4\penalty0 (4):\penalty0 409--423, 1989.

\bibitem[Shahrokhian et~al.(2024)Shahrokhian, Deng, Lin, Ranjan, and Xu]{Shahrokhian2024adaptive}
Anita Shahrokhian, Xinwei Deng, C~Devon Lin, Pritam Ranjan, and Li~Xu.
\newblock Adaptive design for contour estimation from computer experiments with quantitative and qualitative inputs.
\newblock http://arxiv.org/abs/2504.05498, 2024.

\bibitem[Srinivas et~al.(2010)Srinivas, Krause, Kakade, and Seeger]{srinivas2009gaussian}
Niranjan Srinivas, Andreas Krause, Sham~M Kakade, and Matthias Seeger.
\newblock {Gaussian} process optimization in the bandit setting: No regret and experimental design.
\newblock \emph{International Conference on Machine Learning}, 2010.

\bibitem[Xiao et~al.(2021)Xiao, Mandal, Lin, and Deng]{xiao2021ezgp}
Qian Xiao, Abhyuday Mandal, C~Devon Lin, and Xinwei Deng.
\newblock {EzGP}: Easy-to-interpret {Gaussian} process models for computer experiments with both quantitative and qualitative factors.
\newblock \emph{SIAM/ASA Journal on Uncertainty Quantification}, 9\penalty0 (2):\penalty0 333--353, 2021.

\bibitem[Yang et~al.(2020)Yang, Lin, and Ranjan]{yang2020global}
Feng Yang, C~Devon Lin, and Pritam Ranjan.
\newblock Global fitting of the response surface via estimating multiple contours of a simulator.
\newblock \emph{Journal of Statistical Theory and Practice}, 14:\penalty0 1--21, 2020.

\bibitem[Yang et~al.(2023)Yang, Lin, Zhou, and He]{yang2023doubly}
Feng Yang, C~Devon Lin, Yongdao Zhou, and Yuanzhen He.
\newblock Doubly coupled designs for computer experiments with both qualitative and quantitative factors.
\newblock \emph{Statistica Sinica}, 33:\penalty0 1923--1942, 2023.

\bibitem[Yang et~al.(2013)Yang, Lin, Qian, and Lin]{yang2013construction}
Jian-Feng Yang, C~Devon Lin, Peter~ZG Qian, and Dennis~KJ Lin.
\newblock Construction of sliced orthogonal {Latin} hypercube designs.
\newblock \emph{Statistica Sinica}, 23:\penalty0 1117--1130, 2013.

\bibitem[Zhang et~al.(2021)Zhang, Chien, Liu, Xu, and Hong]{zhang2021mixed}
Qiong Zhang, Peter Chien, Qing Liu, Li~Xu, and Yili Hong.
\newblock Mixed-input {Gaussian} process emulators for computer experiments with a large number of categorical levels.
\newblock \emph{Journal of Quality Technology}, 53\penalty0 (4):\penalty0 410--420, 2021.

\bibitem[Zhang et~al.(2020{\natexlab{a}})Zhang, Apley, and Chen]{zhang2020bayesian}
Yichi Zhang, Daniel~W Apley, and Wei Chen.
\newblock Bayesian optimization for materials design with mixed quantitative and qualitative variables.
\newblock \emph{Scientific Reports}, 10\penalty0 (1):\penalty0 1--13, 2020{\natexlab{a}}.

\bibitem[Zhang et~al.(2020{\natexlab{b}})Zhang, Tao, Chen, and Apley]{zhang2020latent}
Yichi Zhang, Siyu Tao, Wei Chen, and Daniel~W Apley.
\newblock A latent variable approach to {Gaussian} process modeling with qualitative and quantitative factors.
\newblock \emph{Technometrics}, 62\penalty0 (3):\penalty0 291--302, 2020{\natexlab{b}}.

\bibitem[Zhou et~al.(2011)Zhou, Qian, and Zhou]{zhou2011simple}
Qiang Zhou, Peter~ZG Qian, and Shiyu Zhou.
\newblock A simple approach to emulation for computer models with qualitative and quantitative factors.
\newblock \emph{Technometrics}, 53\penalty0 (3):\penalty0 266--273, 2011.

\end{thebibliography}

\end{document}